\documentclass[useAMS,usenatbib]{mn2e}
\usepackage{color}
\usepackage{graphicx}
\usepackage{rotating}
\usepackage{lscape}
\usepackage{amsmath,amssymb,amstext}
\usepackage{multirow}
\usepackage{graphicx}
\usepackage{hyperref}

\title[Morphology and metallicity of the Small Magellanic Cloud]
{Morphology and metallicity of the Small Magellanic Cloud using RRab stars}
\author[Deb et al.]{Sukanta Deb$^{1,2}$\thanks{E-mail: sukantodeb@gmail.com}, 
Harinder P. Singh$^2$ , Subhash Kumar$^{1,2}$, Shashi M. 
Kanbur$^3$ \\
$^{1}$Department of Physics, Acharya Narendra Dev College, Govindpuri, Kalkaji, New Delhi 110019, India \\
$^2$Department of Physics \& Astrophysics, University of Delhi,
Delhi 110007, India\\
$^3$State University of New York at Oswego, Oswego, NY 13126, USA \\
}
\begin{document}

\date{Received on ; Accepted on }

\pagerange{\pageref{firstpage}--\pageref{lastpage}} \pubyear{2013}

\maketitle

\label{firstpage}

\begin{abstract}
We present a study of three-dimensional structure of the Small Magellanic
Cloud (SMC). The $V$- and $I$-band light curves of the fundamental mode RR 
Lyrae stars (RRab) obtained by the Optical Gravitational Lensing Experiment
(OGLE)-III project were utilized in order to comprehend the SMC structure. The 
$[Fe/H]-P-\phi_{31}$ relation of \citet{jurc96} is exploited to obtain the 
metallicities. From the three-dimensional RRab distance distributions, 
northeast (NE) arm and main body of the galaxy is identified. Combining 
metallicities with spatial distribution of these tracers, no radial 
metallicity gradient in the SMC has been detected. Dividing the entire sample
into three parts: northeastern (NE), central and southwestern (SW), we find
that the central part has a significantly larger line of sight depth as
compared to rest of the parts, indicating that the SMC may have a bulge.
Results obtained from the $I$-band data seem to be reliable and were further
substantiated using the \citet{smol05} relation. Distribution of SMC RRab
stars were modeled as a tri-axial ellipsoid. Errors in structural parameters
of the SMC ellipsoid were obtained from Monte Carlo simulations. We estimated 
the axes ratios of the galaxy as $1.00\pm 0.000:1.310\pm 0.029:8.269\pm0.934$,
the inclination of the longest axis with line of sight 
$i = 2^{\circ}.265\pm 0^{\circ}.784$, and the position angle of the line of 
nodes $\theta_{\text{lon}}=74^{\circ}.307\pm 0^{\circ}.509$  from the variance 
weighted $I$-band determinations.
\end{abstract}
\begin{keywords}
stars: variables: RR Lyrae-stars:fundamental parameters - stars: Population II - galaxies: statistics - galaxies:structure - galaxies:Magellanic Clouds
\end{keywords}

\section{Introduction}
The Small Magellanic Cloud (SMC) is one of our nearby satellite galaxies. It 
is a gas-rich, dwarf irregular galaxy connected by a hydrogen gas and stellar bridge to the Large Magellanic Cloud (LMC) and located at a distance of 
around $60$ kpc \citep{west97,grac14}. Like LMC, the SMC also exhibits a bar 
which is less pronounced \citep{west97,smit12}. The complexity in the 
structure, dynamics and evolution of the SMC might be attributed to the 
influence of gravitational interactions with the LMC and the Galaxy 
\citep{putm98}. The SMC has a relatively lower heavy metal abundance, lower 
dust content  and significantly larger line of sight depth as compared to the 
LMC and the Galaxy \citep{stan99,stan04,smit12}. With the proliferation of 
accurate and precise variable star data generated from modern astronomical 
surveys and space missions along with their well-established theoretical 
understandings have helped us to gain better knowledge about distances of 
astronomical objects and resolved many issues related to distances and 
distance related parameters such as Galactic and extragalactic structures 
\citep{smit12,kapa12,piet13,grac14}. 
For instance, to gain a better knowledge of the geometric structure of the 
SMC, \citet{grac14} determined the distance to the SMC as $(m-M) = 18.965\pm0.025(\text{stat.})\pm 0.048(\text{sys.})$ mag which corresponds to a distance of $62.1\pm1.9$ kpc using detached, long period eclipsing binaries. Well-detached 
eclipsing binaries serve as a source of accurate and precise distance 
indicators and help in the calibration of the zero point of the cosmic distance 
scale with an accuracy of about $2\%$ \citep{piet13}.

The long-standing and highly controversial questions of the SMC's 
three-dimensional structure and depth along the line of sight have been greatly 
facilitated through the identification and characterization of a statistically 
significant number of different stellar tracers in various large photometric 
surveys. The uniform data sets obtained for a large number of 
different stellar tracers obtained in these surveys in the SMC provide a 
unique opportunity to study the characteristics of the host galaxy to a great 
detail \citep{smit12,kapa12,smit14}.                           

The accurate determination of the geometrical parameters of the SMC plays an 
important role in modeling the interactions between the galaxy, LMC and the 
SMC. In 
this study, the `standard candles' such as the RRab stars have been chosen to 
study the various parameters of the SMC.  RRab stars are the RR Lyrae stars 
which pulsate in the fundamental mode. RR Lyrae stars are 
metal-poor, low-mass and core-helium burning stars that undergo periodic 
radial pulsations \citep{hora04}. They serve as excellent tracers of the 
oldest population of stars in a host galaxy. Their constant mean luminosity 
makes them ideal standard candles \citep{prit11}. They serve as invaluable 
resource which help in exploring the metallicity and structure and hence 
provide substantial insight into the understanding of the formation and 
evolution of the galaxy \citep{pari09,deb14}. They have been used as proxies of the true, 
three-dimensional spatial distribution of host galaxies. Their presence in 
large numbers in different galaxies has led to an improved understanding of 
various geometric parameters of these galaxies. Well-sampled light curves 
covering the entire pulsational period of the RRL stars allow determination of 
their properties with high precision. The derived structural 
parameters extracted from their light curves can be used to determine 
metallicity, interstellar extinction, distance and spatial distribution of the 
stars \citep{piet12}. 

There are a few studies in the literature aimed at determining the geometrical 
structure of the SMC \citep{stan99,crow01,smit12,kapa12,hasc_smc, smit14}. 
Investigating $12$ populous clusters in the SMC, \citet{crow01} determined 
the axes ratios approximately as $1:2:4$ modeling the SMC as a triaxial 
galaxy with declination, right ascension and line of sight depth as the three 
axes. Using the red clump and RR Lyrae stars in the $V$ and $I$-band 
data taken from OGLE-III, \citet{smit12} derived various geometrical parameters
of the SMC.  Their study was based on the relative position of different regions
of the SMC with respect to the SMC mean distance. The parameter determinations 
of the SMC in their work do not quote the errors. It is to be noted that 
the geometry of the SMC is a function of the de-projected Cartesian coordinates 
as well as their errors. Accurate parameter values can only be determined when 
the errors are incorporated in projected Cartesian coordinates along with the 
reliable distance determinations. Also, the metallicity determination of the RR 
Lyrae stars were not attempted in their work which is very crucial in 
understanding of substructure of the galaxy. The axes ratios, inclination angle 
of the longest axes along the line of sight ($i$) and position angle 
of the line of nodes ($\theta_{lon}$) obtained by \citet{smit12} from the 
distribution of the SMC RR Lyraes were found to be 
$1:1.33:6.47$, $0^{\circ}.4$ and $74^{\circ}.4$, 
respectively from the entire dataset. Taking different coverage of the dataset 
and removing the isolated northwestern fields, different structural parameters 
of the SMC ellipsoid were also attempted and estimated by \citet{smit12}.
On the other hand, using the $454$ OGLE-III RRab data in the $V$-band, 
\citet{kapa12} used the line of sight distances and the metallicity values 
to investigate the possible presence of different structures containing 
different populations in the SMC. Using spherical shells of different radii of 
$2.5$, $3.0$ and $3.5$ kpc, they obtained the axes ratios of the SMC ellipsoid 
as: $1:1.21:1.57$, $1:1.18:1.53$ and $1:1.23:1.80$, respectively, whereas, 
inclination angle and position angle of line of nodes of the SMC were not 
calculated in their analyses. From an analysis of radial velocities of $150$ 
carbon stars, \citet{kunk00} estimated the inclination of the SMC orbital 
plane to be $i = 73^{\circ}\pm4^{\circ}$ relative to the plane of the sky. 
Using Cepheids, \citet{cald86} and \citet{groe00} estimated $i$ as 
$70^{\circ}\pm3^{\circ}$ and $68^{\circ}\pm2^{\circ}$, respectively. The 
position angle of line of nodes was estimated to be $\sim~148^{\circ}$ 
\citep{groe00}. Estimates of inclination angle and position angle of line of 
nodes obtained by \citet{hasc_smc} from the $I$-band OGLE III SMC 
RRab data are $7^{\circ}\pm15^{\circ}$ and $83^{\circ}\pm 21^{\circ}$, 
respectively. From the study of distance distribution of Cepheids in 
the SMC using the OGLE-III data, \citet{hasc_smc} found that these young 
populations are differently oriented than the old population RR Lyrae stars.
The inclination angle and position angle parameters of the SMC obtained from 
Cepheids by \citet{hasc_smc} are $74^{\circ}\pm9^{\circ}$ and 
$66^{\circ}\pm15^{\circ}$, respectively. Other recent estimates of inclination 
angle and position angle of lines of nodes for the SMC obtained from Cepheids
are $64^{\circ}.4\pm0^{\circ}.7$ and $155^{\circ}.3\pm 6^{\circ}.3$, 
respectively \citep{smit14}.

In this paper, we use available $V$ and $I$-band OGLE-III data in order 
to determine the absolute magnitudes and metallicities of RRab stars. These
measurements along with equatorial coordinates, viz., right ascension 
($\alpha$) and declination ($\delta$) have been utilized to get an insight into 
the understanding of the three-dimensional structure of the SMC and its 
metallicity distribution. The availability of data in two different bands  
and the use of different methodologies of distance determinations 
provide a unique opportunity to compare and contrast various structural 
parameter determinations. The roadmap of the present investigation is to use 
these tracers for the distance determinations and obtain distances relative to 
the center of the SMC. The principal axis transformation method as described in 
\citet{deb14} was used to get the viewing angle and geometrical parameters of 
the galaxy. Section \ref{data} describes the data used for the analysis in the 
present study. Fourier decomposition method and sample selection criteria are 
described in section~\ref{fd}. Section \ref{extinction} focusses on extinction 
measurements for the OGLE-III data along the line of sight using 
the SMC extinction map from \citet{zari99,zari02}. Determination of 
metallicities \citep{jurc96} and absolute magnitudes \citep{kova01,cate08} 
from $V$-band and $I$-band data calibrated to the $V$-band
are described in section~\ref{metallicity}. Viewing angle parameters and 
geometrical parameters of the SMC determined for the above dataset are 
described in section~\ref{structure}. Line of sight depth and existence of 
metallicity gradient are examined in sections~\ref{los} and \ref{metgrad}, 
respectively. Results obtained from the above analyses were further 
substantiated as described in section~\ref{smol} with \citet{smol05} 
metallicity relation using the $I$-band light curve data and exploiting the 
$M_{V}-[Fe/H]$ relation of \citet{cate08}. Exponential disk model and 
\citet{king62} three parameter model for the radial number density 
distribution of RRab stars in the $I$-band are discussed in 
section~\ref{scale}. Lastly, summary and discussions of the present 
investigation are laid down in section~\ref{summary}.         
\section{The Data}
\label{data}
RRab stars for the present study were selected from OGLE-III catalog of 
variable stars that consists of  $8$-year archival data identified and 
characterized by the Fourier coefficients of the light curves \citep{sosz09}. 
The catalog comprises $1933$ RRab stars in the $I$-band having a mean period of 
$<P_{ab}> = 0.596$ days with typical accuracy of $\sigma_{P}/{P} \simeq 
10^{-5}$ \citep{sosz09}. The catalog is based on the 
observations carried out at the Las campanas Observatory, Chile with a $1.3$-m 
telescope. The time series observations of the SMC consist of $800$ 
nights between June $2001$ and May $2009$. The OGLE field in the SMC 
covers nearly $16$ $\deg^{2}$ covering the bar and the wing. Most of the 
observations were carried out using the Cousins $I$-band filter with exposure 
time of $180$ s having an average of $400$ photometric observations. The 
catalog also contains $V$-band light curves of $1887$ stars having an average 
of $30$ data points per light curve with the
exposure time of $225$ s \citep{sosz09}.
\section{Fourier decomposition method and sample selection}
\label{fd}
\begin{figure*}
\begin{center}
\includegraphics[width=0.8\textwidth,keepaspectratio]{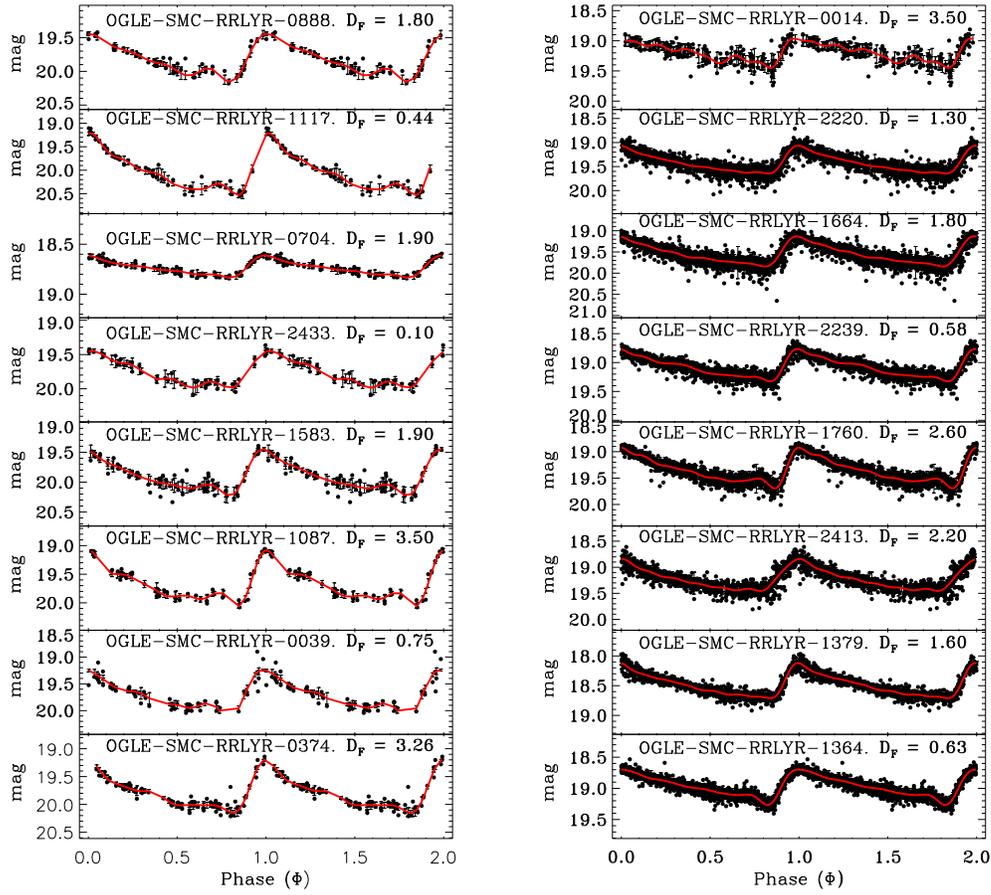}
\caption{A collection of randomly selected sample of RRab light curves with 
their corresponding OGLE IDs and deviation parameters ($D_{F}$) are shown. 
Left panel shows the phased light curve data in $V$-band while the right panel 
depicts the same in the $I$-band.} 
\label{lcplot}
\end{center}
\end{figure*}

The Fourier decomposition method was used to obtain $V$ and $I$-band light 
curve parameters of the SMC RRab stars. The light curves were fitted with a 
Fourier sine series of the form \citep{deb09,deb10,deb14}
\begin{equation}
\label{fdtech}
m(t) = A_{0}+\sum_{i=1}^{N} A_{i} \sin\left[i\omega (t-t_{0}) +  \phi_{ i}\right],
\end{equation} where $m(t)$ is the observed magnitude, $A_{0}$ is the mean magnitude, $\omega$=2$\pi/P$ is the angular frequency, $P$ is the pulsational 
period of the star in days and $t$ is the time of observation. $t_{0}$ represents the epoch of maximum light 
(either in the $V$ or the $I$-band) and is used to obtain a phased light curve 
which has maximum light at phase zero. 
$A_{i}$'s and $\phi_{i}$'s are the $i$th order Fourier coefficients 
and $N$ is the order of the fit. Eqn.~\ref{fdtech} has $2N+1$ unknown 
parameters. To solve for these parameters, at least the same number of data 
points are required. The light curve phase was obtained using
\begin{displaymath}
\Phi =\frac{\left( t-t_{0}\right) }{P}-Int\left( \frac{\left(
t-t_{0}\right) }{P}\right).
\end{displaymath}  
Here $\Phi \in [0,1]$ represents one pulsation cycle of the RRab stars.
The pulsation periods ($P$) and the epoch ($t_{0}$) are taken from the OGLE 
catalog. The Fourier parameter code as described in \citet{deb10} was used 
to obtain various Fourier parameters on the right hand side of 
Eqn.~\ref{fdtech}. Fifth order and seventh order Fourier fits 
were employed to model the RRab  $V$ and $I$-band light curves, respectively. 
The Fourier fitted $V$ and $I$-band light curves of a sample of RRab stars 
are shown in Fig.~\ref{lcplot}. The phase differences, 
$\phi_{i1} = \phi_{i}-i\phi_{1}$ and amplitude ratios, 
$R_{i1} = (A_{i}/A_{1})$, $i > 1$ were evaluated and standard errors were determined using the formulae of \citet{deb10}. A clean sample of RRab stars for the 
present analysis were selected with the criteria based on OGLE-determined 
periods, the mean magnitude ($A_{0}$) and the peak-to-peak $V$ and $I$-band 
amplitude ($A_{V},A_{I}$) determined from the Fourier analysis of the phased 
light curves. RRab stars with periods $P \ge 0.4$ days, mean magnitude 
$A_{0} \ge 18.0$ mag ($I$-band)and amplitude $0.1 \le A_{I} \le 1.2$ mag were 
chosen for the $I$-band light curves. On the other hand, to select the 
$V$-band light curves, the same criteria on periods has been applied. While the 
following criteria were applied on $A_{0}$ and amplitude: $A_{0} \ge 18.5$ mag 
($V$-band) and amplitude $0.2 \le A_{V} \le 1.5$. These selection criteria 
were applied in order to collect a clean sample of RRab stars of the SMC which 
are free from any contamination due to the foreground objects of our Galaxy 
and the LMC. The application of these selection criteria along with further 
application of compatibility test of \citet{jurc96} reduces 
the number of RRab stars in the $V$ and $I$-band light curves to $1142$ and 
$1543$, respectively for the physical parameter estimation and determination 
of structure of the SMC. The compatibility test of \citet{jurc96} ensures to 
reject those RRab stars that have deviation parameter values $D_{F} > 5$.  
\section{Interstellar Extinction}
\label{extinction}
\begin{figure*}
\vspace{0.02\linewidth}
\begin{tabular}{cc}
\vspace{+0.01\linewidth}
  \resizebox{0.50\linewidth}{!}{\includegraphics*{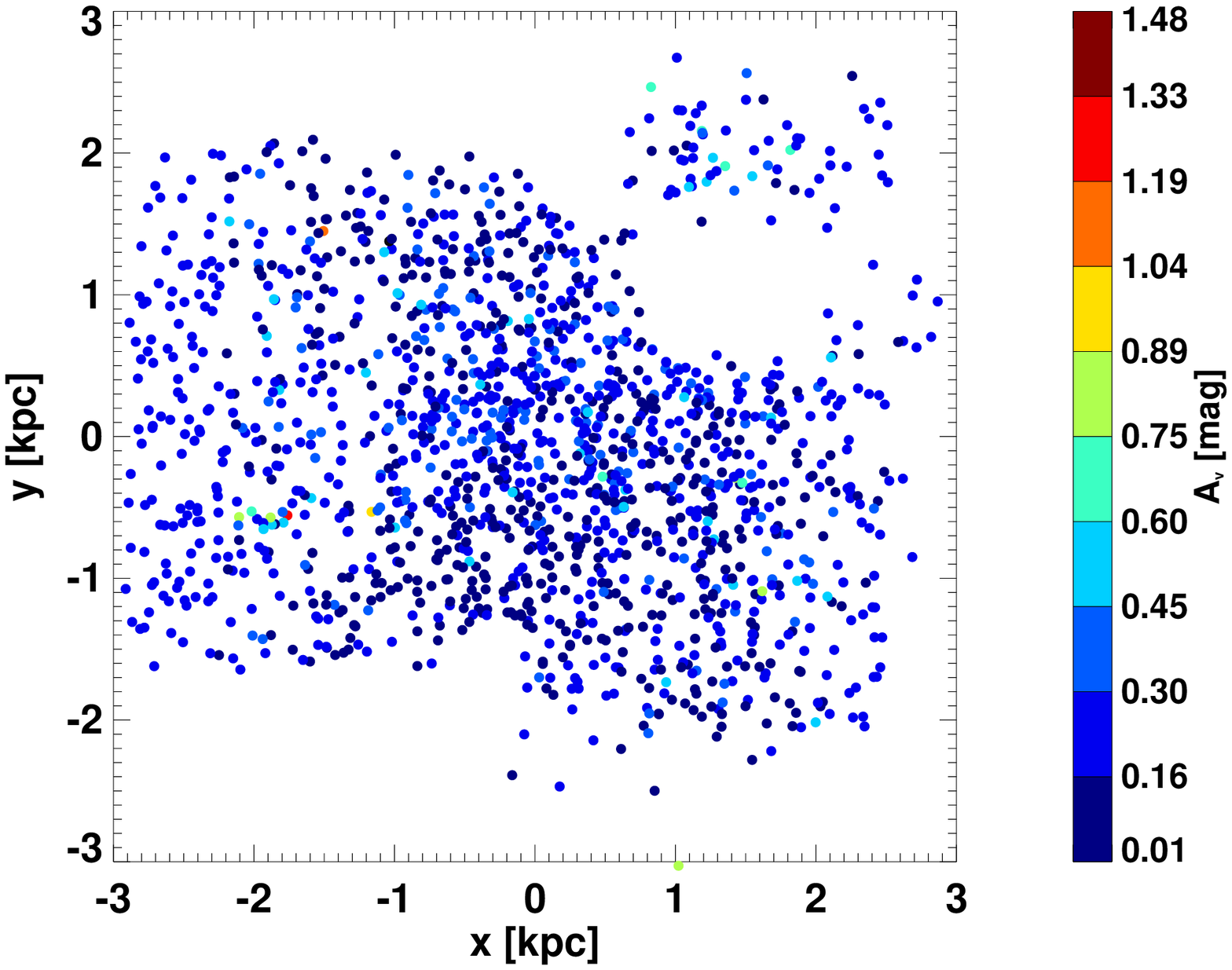}}&
  \resizebox{0.50\linewidth}{!}{\includegraphics*{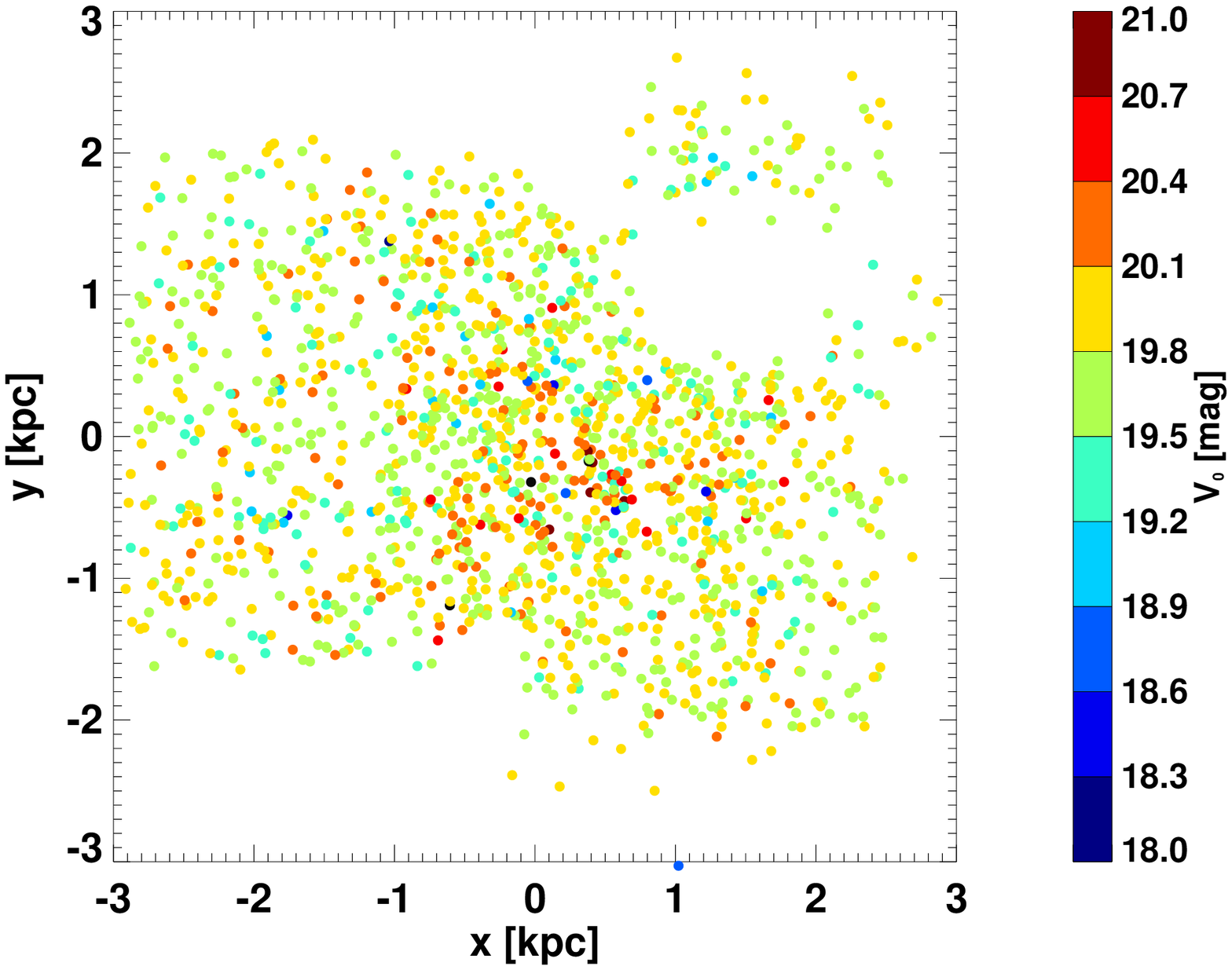}}\\
\vspace{-0.04\linewidth}
\end{tabular}
\caption{Two-dimensional color bar plot of the interstellar extinction values 
($A_{V}$) and extinction corrected mean magnitudes ($V_{0}$) of RRab stars.}
\label{cbar}
\end{figure*}

Interstellar medium plays an important role in dimming and reddening of the 
star light. In order to determine accurate distances to stars, interstellar 
reddening must be accounted for. Various methods to determine interstellar 
extinction exist in the literature using different tracers. Observations of 
multiband photometry available for RR Lyrae stars serve to resolve internal 
extinction in the case of host galaxy \citep{pejc09}. For the determination of 
interstellar extinction in the case of individual RRab stars in this paper, we adopted the \citet[hereafter, Zaritsky map]{zari99,zari02} extinction map for 
the SMC which comes in FITS  (Flexible Image Transport System) format. The SMC 
extinction map for cool stars is downloaded from \footnote{\url{http://djuma.as.arizona.edu/~dennis/mcsurvey/Data_Products.html}}, since the cool star data are 
characteristics of the RR Lyrae with $T_{\rm eff}\sim 6500$ K. The Magellanic 
Cloud Photometric Survey (MCPS) contains $UBVI$ photometry and extinction map 
for the central $18$ deg$^{2}$  area of the SMC \citep{zari02}. For a given 
RRab location, the map returns the $V$-band extinction value, $A_{V}$. The 
FITS image of the Zaritsky map containing the extinction values ($A_{v}$) 
is an array of $(240\times280)$ pixels with reference pixel coordinate as 
$(X,Y)=(120,140)$ corresponding to the 
$(RA,DEC) = (12^{\circ}.750, -72^{\circ}.699)$.  The $RA$ and $DEC$ pixel 
scales in the image are $\sim 0.0167^{\circ}$/pixel which correspond to 
$1^{\prime}$/pixel. The IRAF task `imhead' was used to get the details of the 
FITS image header. The number of stars used to derive extinction for 
map pixels were at least $3$ where the median extinction value was adopted 
\citep{zari99}. The Zaritsky map  has regions of high and low spatial resolution but relatively uniform signal-to-noise ratio. The details of smoothing 
algorithm applied to derive the extinction values for the map pixels are 
outlined in \citet{zari99}. The following 
procedure was adopted to get the extinction values of the sample of 
RRab stars. RA$(\alpha)$ and Dec$(\delta)$ of the sample stars were first 
converted into $X/Y$ pixel coordinates using $sky2xy$ followed by running 
$getpix$ on the FITS image to get interstellar extinction ($A_{V}$) at that 
$X/Y$ position. Both $sky2xy$ and $getpix$ are the codes from WCSTOOLS 
\footnote{\url{http://tdc-www.harvard.edu/wcstools/}}. For a few stars, we 
get a null value of $A_{V}$. For these stars, the mean value of $A_{V}=0.155$ 
mag appropriate for the SMC was used. The average value of $A_{V}$ for the 
OGLE-III RRab's is $0.155\pm0.127$ mag which translates into a mean $E(B-V)$ 
of $0.048\pm0.039$ mag. This is comparable to the mean $E(B-V)$ value of 
$0.054$ mag and $E(V-I)$ value of $0.07\pm0.06$ mag, respectively by 
\citet{cald85} obtained from the analysis of $48$ Cepheids and \citet{hasc12} 
from the analysis of $1529$ SMC RRab stars.  Since no formal errors are given 
by the map for the individual stars, an error margin of $20\%$  is assumed in 
the individual extinction values if these estimates from the map are taken to 
be reliable. The extinction maps derived from the OGLE RRab stars by 
\citet[hereafter, Haschke map]{hasc12}  were found to be in good agreement 
with those obtained from the Zaritsky reddening map. Fig.~\ref{cbar} depicts 
two-dimensional color bar plot of interstelar 
extinction values ($A_{V}$) and extinction corrected mean magnitudes 
($V_{0}$) of RRab stars. $x$ and $y$ in the figure denote projected 
Cartesian coordinates. 
\section{Metallicities and Absolute Magnitudes}
\label{metallicity}
Metallicities of a large number of stars found in a certain galaxy play 
pivotal role in understanding the formation of various structures of that 
galaxy. Assembly of stars having certain metallicity ranges concentrated in a 
particular part of a galaxy tells the epoch of its formation. Although the 
metallicity is an important parameter for understanding chemical properties of 
the underlying stars and hence that of the distant host galaxy 
and its formation, it is hard and time-consuming to obtain the metallicity
spectroscopically for a large number of stars. Calibrated relationships 
between various photometric light curve parameters and spectroscopic 
metallicities of RRab stars provide efficient and robust ways for determining 
the metallicities in various globular clusters, the Galaxy, the SMC and 
the LMC by recent ground-based automated surveys. Previous studies have shown 
that metallicities of RRab stars can be derived from an empirical relation 
connecting spectroscopic metallicities and the light curve parameters as 
proxies \citep{simo88,jurc96,sand04,neme13}.  These studies have shown that 
metalicities of RRab stars can be linked the period with some intermediate 
parameters such as amplitude $(A)$, Fourier phase  parameters $\phi_{21}$, 
$\phi_{31}$ \citep{jurc96,alco00,sand04,neme13}. 
The correlation between the light curve structures  of RR Lyrae stars such as 
the period ($P$), Fourier coefficients ($\phi_{21}, \phi_{31}$) and their 
metallicities was first noticed and studied in the pioneering work by 
\citet{simo88}. This work has motivated many investigators such as 
\citet{jurc96} and his collaborators, \citet{sand04} to derive semi-empirical 
relations involving light curve structures and spectroscopically determined 
metallicities. 
\begin{figure*}
\vspace{0.02\linewidth}
\begin{tabular}{cc}
\vspace{+0.01\linewidth}
  \resizebox{0.46\linewidth}{!}{\includegraphics*{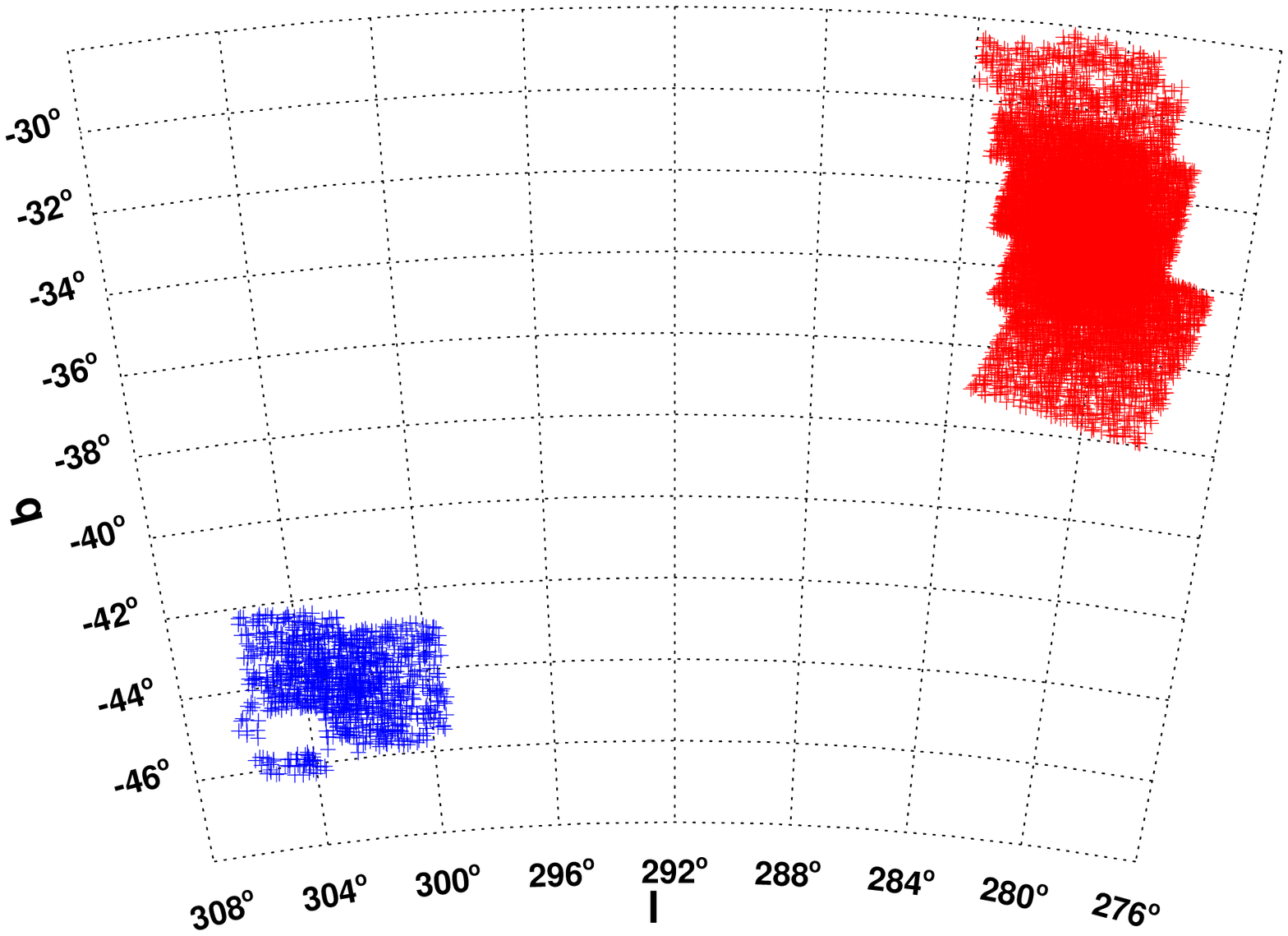}}&
  \resizebox{0.46\linewidth}{!}{\includegraphics*{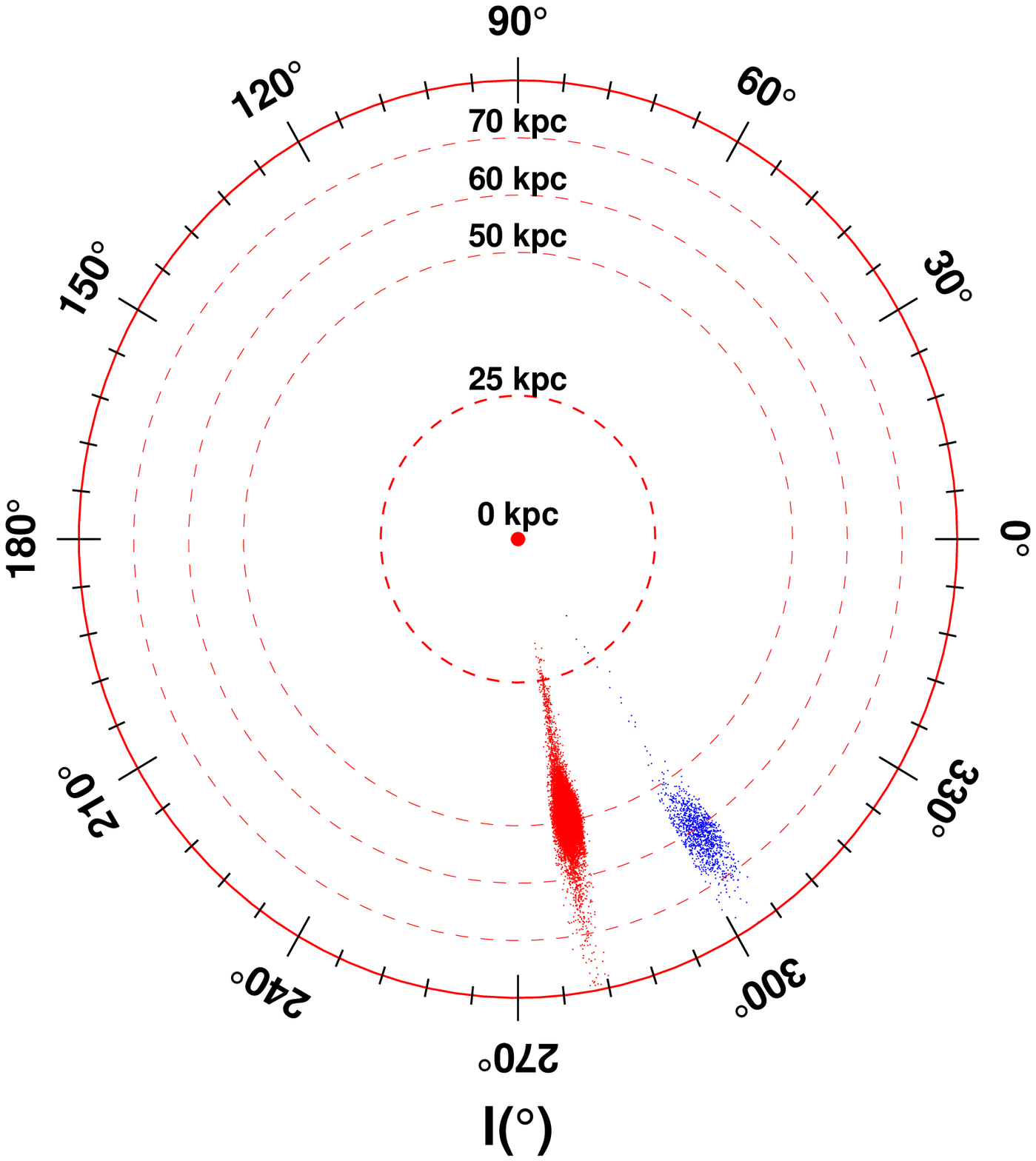}}\\
\vspace{-0.04\linewidth}
\end{tabular}
\caption{Left panel shows LMC and SMC RRab stars in 
galactic coordinates $(l,b)$. Right panel depicts polar plot of distance 
distribution of LMC and SMC RRab stars in galactic coordinates ($l,b$). The 
dashed lines are set at distances $D = 25, 50, 60, 70~\text{kpc}$, 
respectively. LMC and SMC stars are shown as red and blue data points, 
respectively. LMC RRab stars are taken from \citet{deb14}. The tails of 
distance distributions of the two Magellanic Clouds account for possible
contaminations from the foreground and background RRab stars.}
\label{polar_plot}
\end{figure*}

The unprecedented amount of photometric light curve data generated from the 
recent automated surveys such as OGLE, ASAS (All Sky Automated Survey), 
and highly accurate and precise stellar data obtained from the NASA's $Kepler$ 
mission have provided a new avenue into the understanding of various 
astrophysical problems more accurately. Although the $[Fe/H]-P-\phi_{31}$ 
relation of \citet{jurc96} is widely used in the determination of 
metallcities of RR Lyrae stars from their photometric light curves, 
accurate light curve data of these stars generated from the NASA's $Kepler$ 
mission have helped in refining the model parameters of existing 
$[Fe/H]-P-\phi_{31}$ relation and calibrate new relations based on the new 
data \citep{neme13}. These relations will prove to improve our knowledge in 
understanding the chemical history of the stellar tracers as well as structure 
of the host galaxy from their distance distributions. Metallicities 
$[Fe/H]_{\text {spec}}$ of the corresponding light curves in \citet{neme13} 
were obtained from the high resolution spectroscopic measurements.The 
empirical relation of \citet{jurc96} connecting $\phi_{31}$ in the $V$- band, 
period $P$ and the metallicity $[Fe/H]$ is given by
\begin{equation}
\label{jk96}
[Fe/H]_{JK} = -5.038 - 5.394\, P + 1.345\,\phi_{31},~~\sigma = 0.14.
\end{equation}
On the other hand, the non-linear relation connecting $P-\phi_{31}-[Fe/H]$ 
using $26$ RRab stars obtained by the NASA's $Kepler$ space telescope in the 
$K_{p}$-band is given by \citep{neme13}  
\begin{eqnarray}
[Fe/H]_{\text{N13}}=(-8.65\pm4.64)+(-40.12\pm5.18) P+~~~\nonumber \\ 
(5.97\pm1.72)P \phi_{31}^{s}+(6.27\pm0.96)(\phi_{31}^{s})^2,~~~~\sigma=0.084
\end{eqnarray}
where $\phi_{31}^{s}$ is the mean $\phi_{31}^{s}(Kp)$ value and $[Fe/H]$ is 
the spectroscopic metallicity value obtained from the recent new high 
resolution spectroscopic measurements \citep{neme13}. In order to apply the 
N13 relation to the $V$-band data, \citet{neme11} gave the following 
inter-relation 
\begin{equation}
\phi_{31}^{s}(V)=\phi_{31}^{s}(Kp)-(0.151\pm0.026). 
\end{equation}   
The above inter-relation was derived by \citet{neme13} using the ASAS $V$-band 
and the Kepler photometric light curve based on only $3$ stars. The 
inter-relation was recently refined using  $V$ band photometric data of $34$
common stars in the Kepler field and the following relation was obtained 
\citep{jeon14}
\begin{equation}  
\phi_{31}^{s}(V)=\phi_{31}^{s}(Kp)-(0.174\pm0.085).
\end{equation} 
In the present analysis, we have determined metallicities using \citet{jurc96} 
linear 
relation. Although the \citet{neme13} relation is based on the 
highly accurate and precise $Kepler$ data, we have found from numerical 
analysis that formal errors on the calculated $[Fe/H]$ values from the derived 
relation turn out to be very large using the propagation of error formula 
of \citet{bevi03}. This might be attributed to the larger errors in $\phi_{31}$ 
values for the OGLE-III data and more number of complexity parameters of the 
derived relation. Hence because of limitations of the relation restricted to 
highly accurate dataset, we  have to leave out using the \citet{neme13} 
relation for the OGLE-III dataset. The absolute magnitudes are determined 
using the following relation \citep{cate08}
\begin{equation}        
\label{cate_eq}
M_{V} = (0.23\pm0.04)[Fe/H]+(0.984\pm0.127),
\end{equation}
where $[Fe/H]$ is the metallicity in the \citet{zinn84} scale. The metallicity 
values obtained from Eqn.~\ref{jk96} are in \citet{jurc96} scale, which can be 
transformed into the metallicity scale of \citet{zinn84} using the relation 
from \citet{jurc95}:
\begin{equation}         
\label{zw_scale}
[Fe/H]=\frac{[Fe/H]_{JK}-0.88}{1.431}.
\end{equation}
 
\begin{figure*}
\vspace{0.02\linewidth}
\begin{tabular}{cc}
\vspace{+0.01\linewidth}
  \resizebox{0.47\linewidth}{!}{\includegraphics*{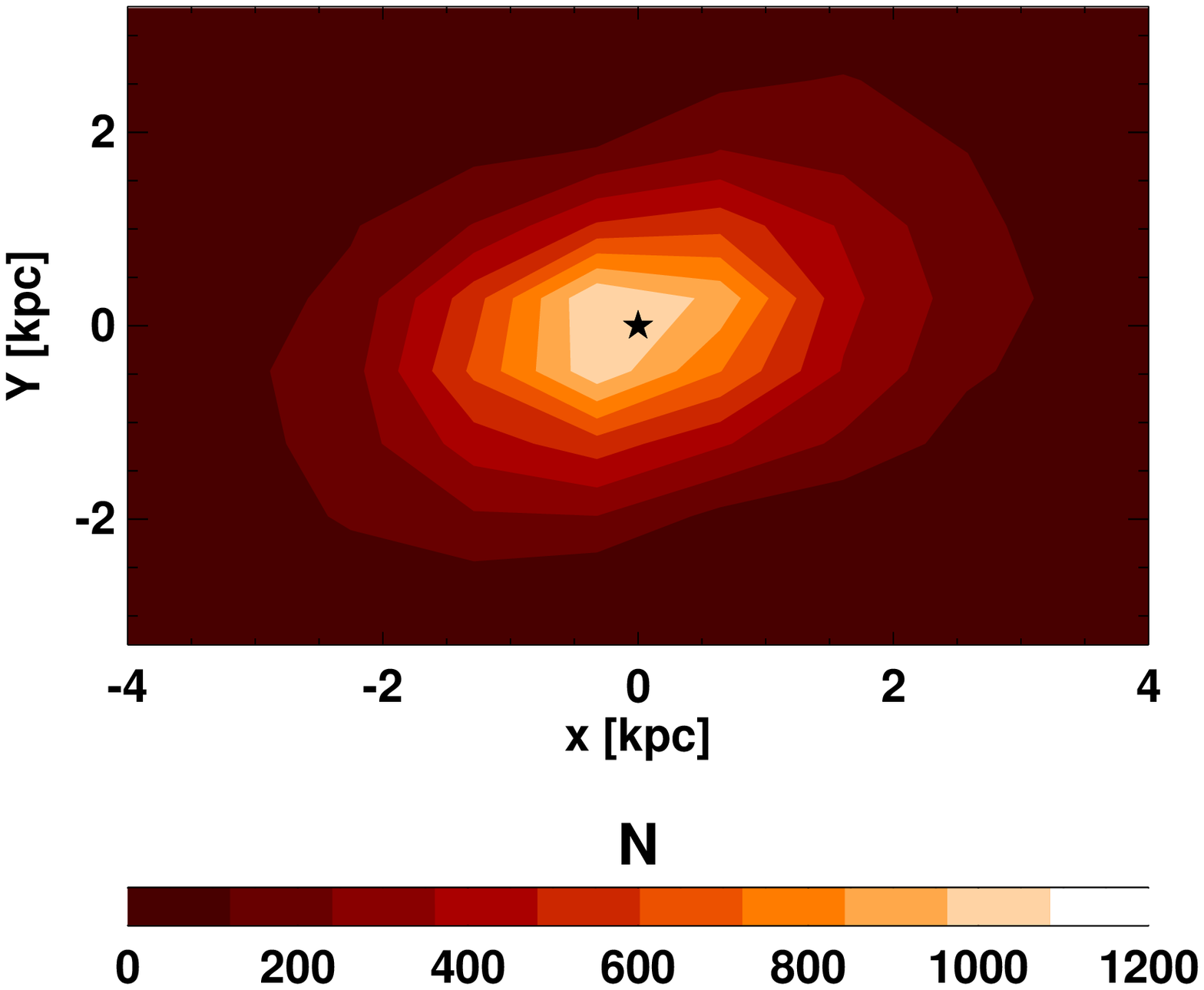}}&
  \resizebox{0.47\linewidth}{!}{\includegraphics*{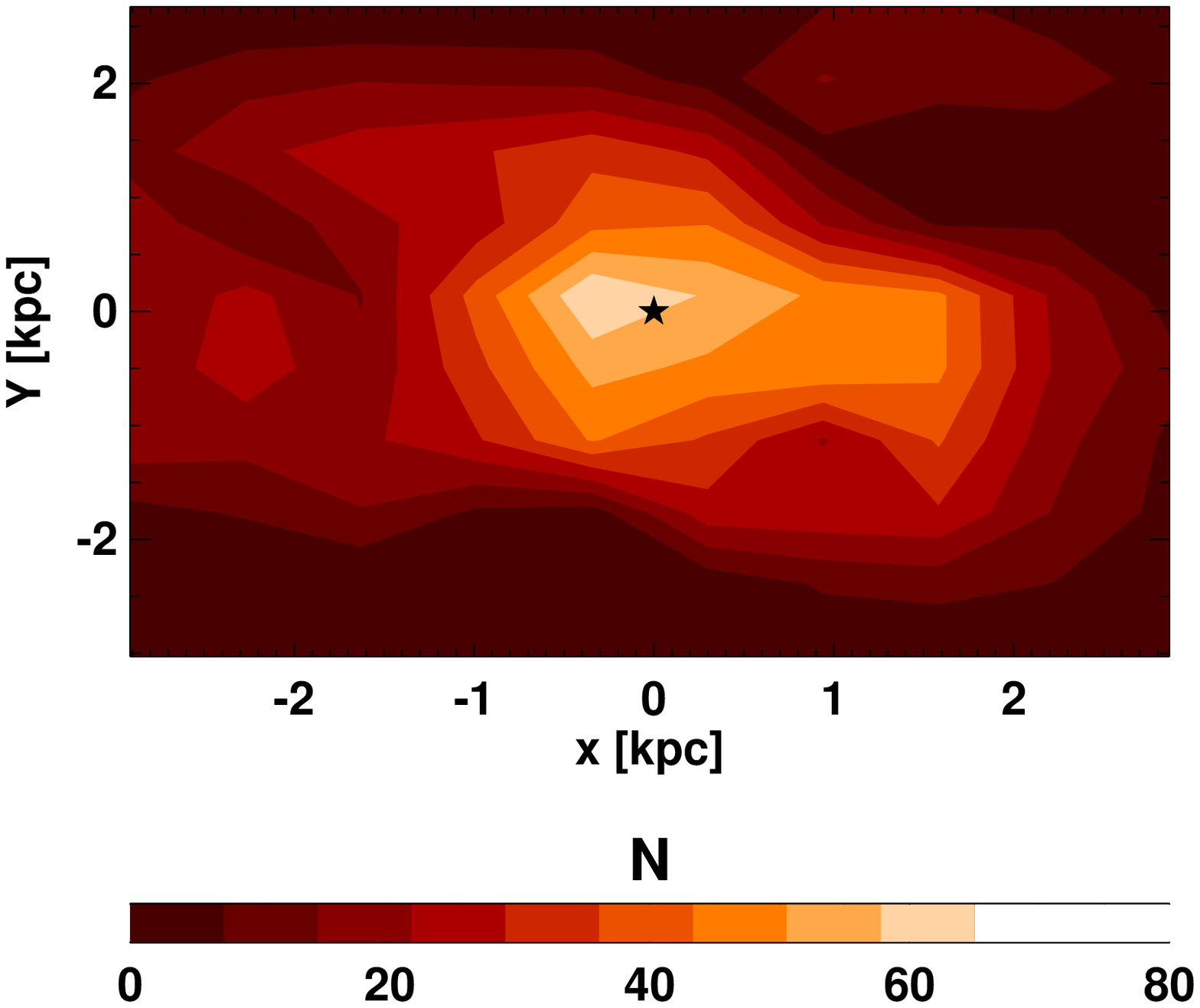}}\\
\vspace{-0.04\linewidth}
\end{tabular}
\caption{Left panel shows two-dimensional density contours of the LMC RRab 
stars taken from \citet{deb14}. Right panel shows the same for the SMC RRab 
stars in the present study. The star symbol denotes the location of the 
centroid.}  
\label{density_plot}
\end{figure*}

\begin{figure}
\begin{center}
\includegraphics[width=0.48\textwidth,keepaspectratio]{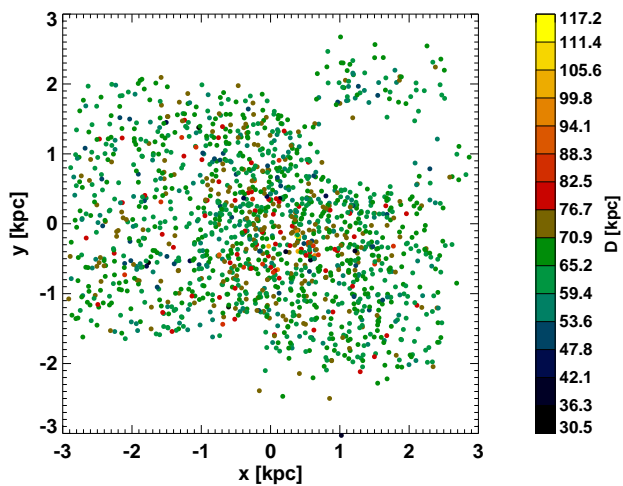}
\caption{Two-dimensional color bar plot of distance distribution of SMC 
RRab stars as a funtion of $(x,y)$ coordinates.}
\label{col_dist}
\end{center}
\end{figure}

Determination of absolute magnitudes of the present sample of RRab stars using
Eqn.~\ref{cate_eq} is denoted as `MVC'. The absolute magnitudes of 
the selected samples of RRab stars were also determined using the following 
empirical relation in terms of the period ($P$) and the Fourier coefficients 
($A_{1}$ and $A_{3}$) \citep{kova01,ferr10}:
\begin{equation}
\label{abs_mag}
M_{V}=-1.876\log{P}-1.158A_{1}+0.821A_{3}+0.41.
\end{equation}
Absolute magnitude determination using this relation is referred to as 
`MVF'. Both the  methods of absolute magnitude determination were applied 
in $V$ and $I$-band to calculate distances of the present sample of RRab stars. 
Suitable inter-relations between the Fourier parameters  in the $V$ and 
$I$-band were utilized as discussed in \citet{deb10,deb14}. Geometrical 
parameter determinations of the SMC using two different relations in two 
different bands carried out in this paper will help us in improving their 
estimates. Fig.~\ref{polar_plot} shows polar plot of distance distribution of 
the SMC RRab stars ($I$-band calibrated to $V$-band) in galactic ($l,b$) 
coordinates. LMC RRab stars as in \citet{deb14} are also overplotted. The 
two-dimensional density contours of the distribution of the LMC RRab stars 
taken from \citet{deb14} and the SMC RRab stars in the present study are shown 
in Fig.~\ref{density_plot}. The star symbol denotes the location of the 
centroid of the present sample. Two-dimensional color bar plot of distance 
distributions of SMC RRab stars is shown in Fig.~\ref{col_dist}.        
\section{Structure of the SMC}
\label{structure}
\begin{figure*}
\vspace{0.02\linewidth}
\begin{tabular}{cc}
\vspace{+0.01\linewidth}
  \resizebox{0.46\linewidth}{!}{\includegraphics*{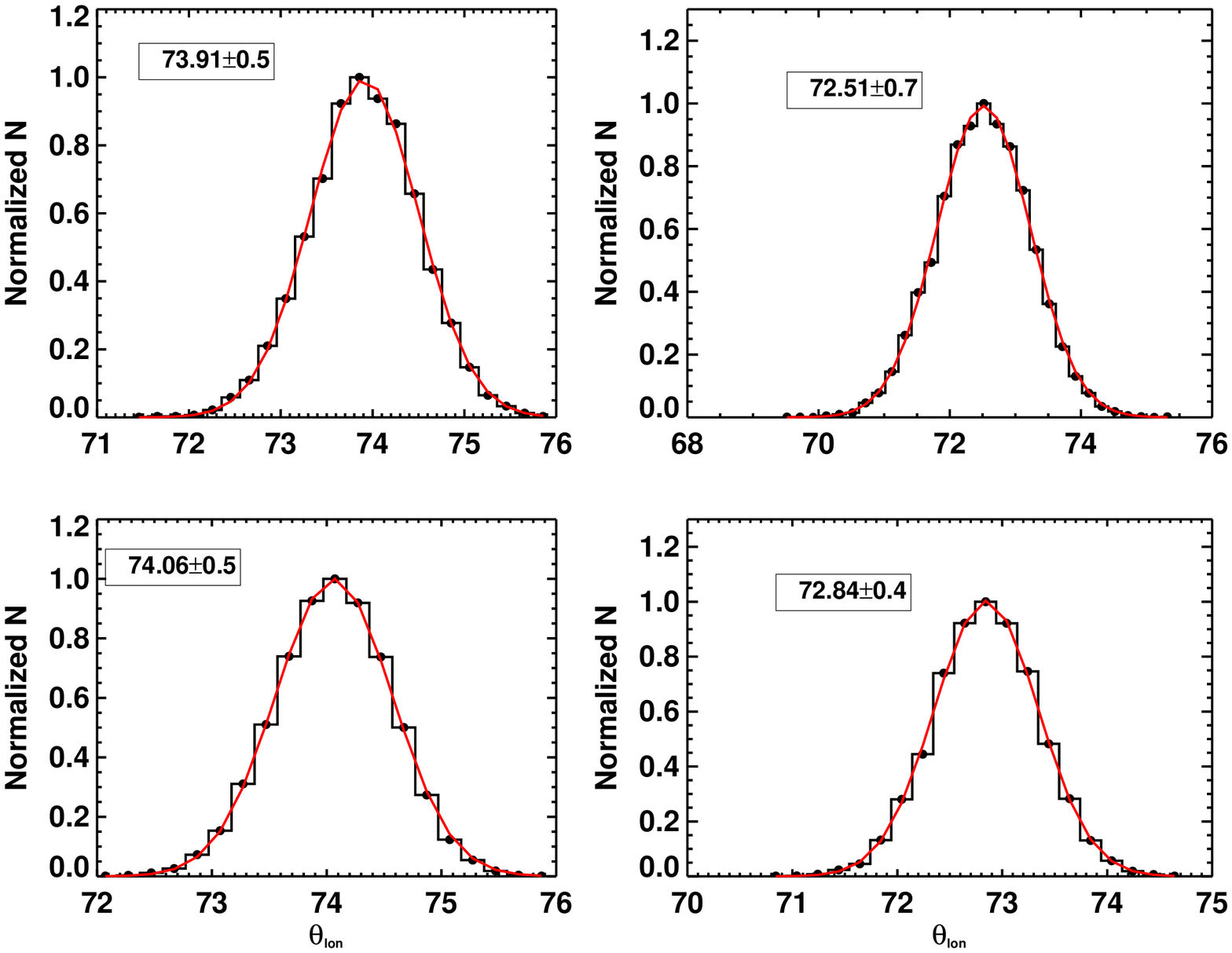}}&
  \resizebox{0.46\linewidth}{!}{\includegraphics*{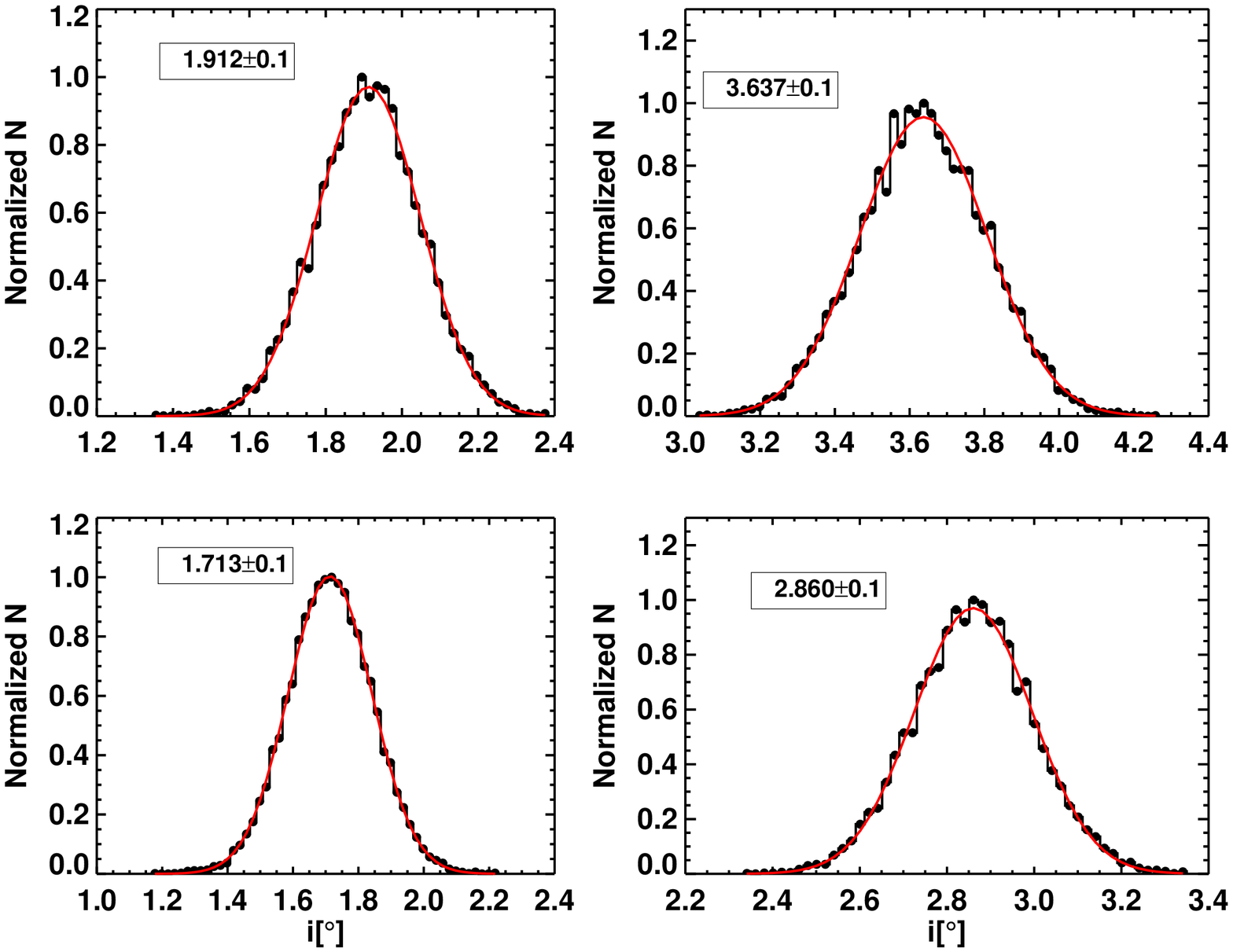}}\\
\vspace{+0.01\linewidth}
  \resizebox{0.46\linewidth}{!}{\includegraphics*{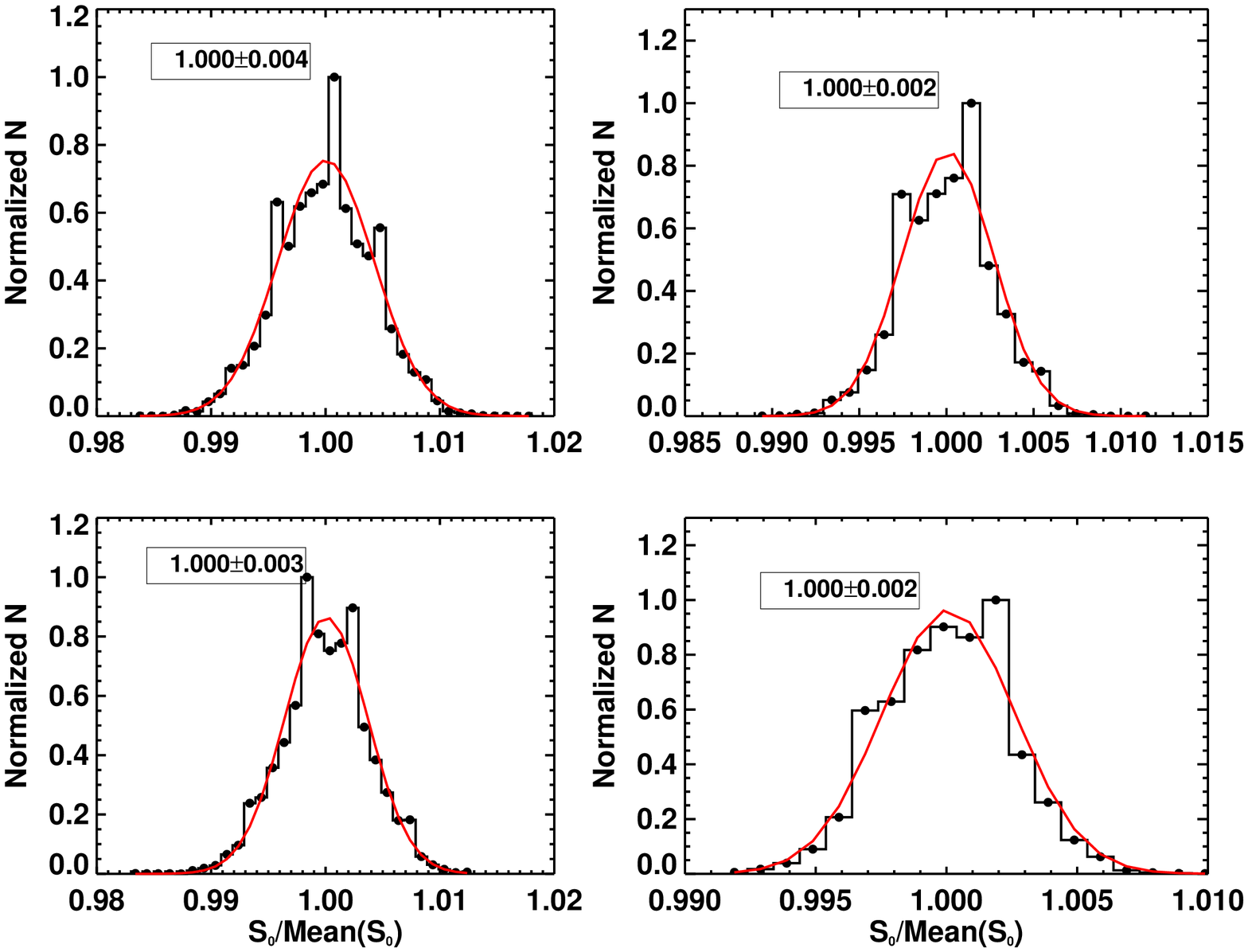}}&
  \resizebox{0.46\linewidth}{!}{\includegraphics*{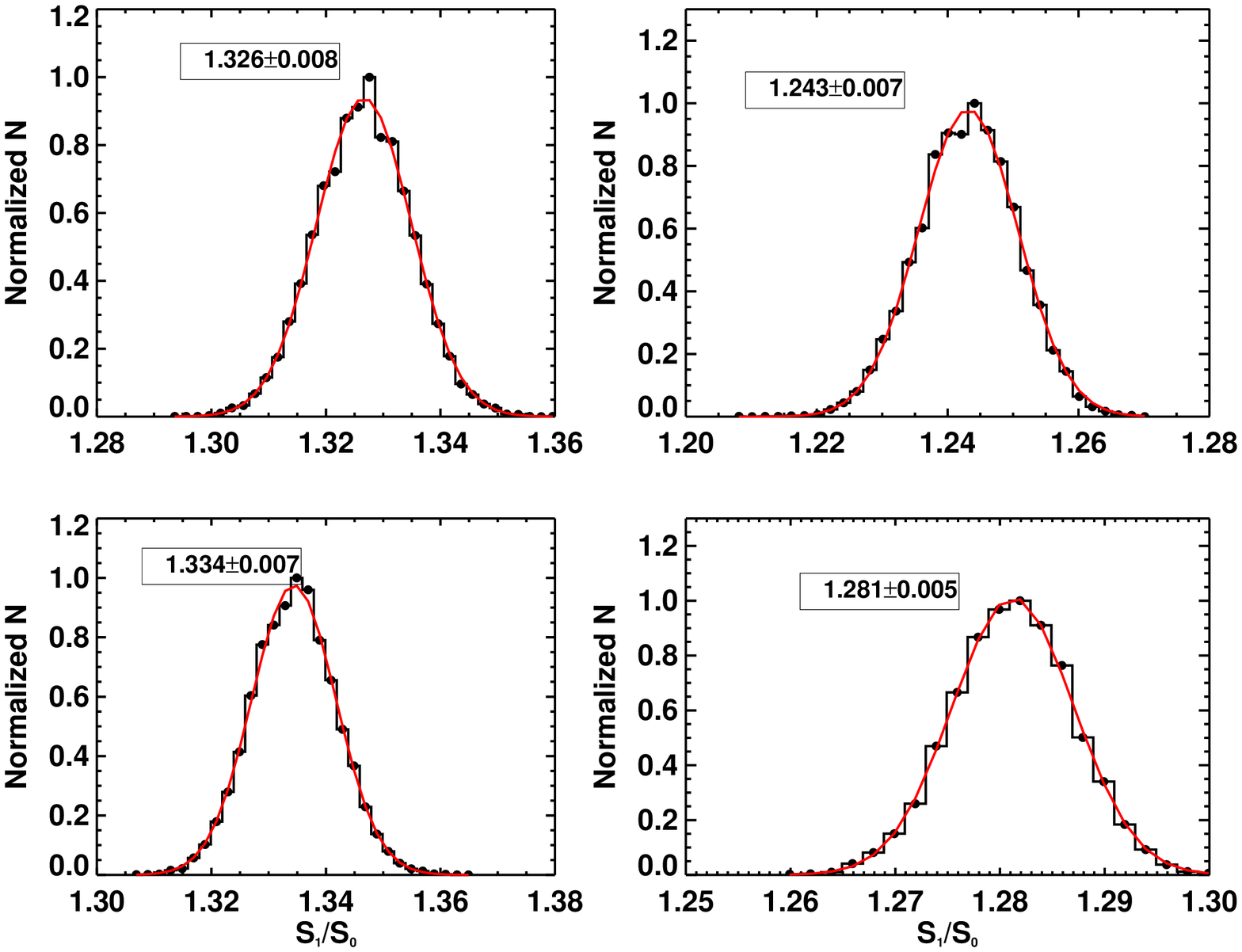}}\\
\vspace{-0.04\linewidth}
\end{tabular}
\caption{Distribution of viewing angles and geometrical parameters 
of the SMC in histograms. Solid red lines denote Gaussian fit to the resulting 
distributions.}
\label{fig_all}
\end{figure*}

\begin{figure}
\includegraphics[width=0.46\textwidth,keepaspectratio]{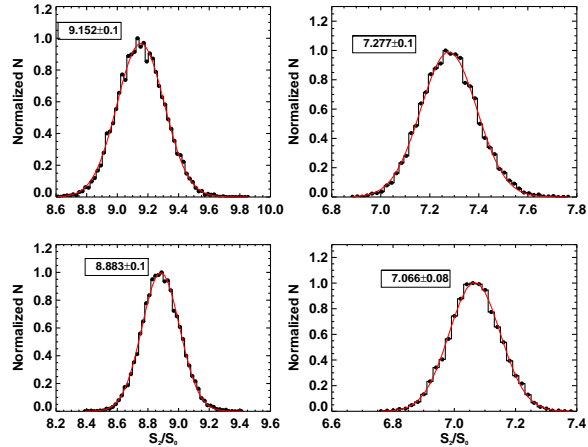}
\caption{Distribution of $S_{2}/S_{0}$ ratio  for $10^{5}$ Monte Carlo
iterations in histograms. Solid red lines denote Gaussian fit to the resulting
distributions.}
\label{axis}
\end{figure}

Availability of a statistically significant sample of RRab stars from 
the OGLE-III data in $I$ and $V$-band provides an opportunity to compare
and contrast the two different bands to geometric parameter determination of 
the SMC. Furthermore, structural parameter determinations of the SMC 
substantiated using two independent distance determination formula will also 
help us in improving the structural parameters of the SMC. In order to 
determine the structure, we have converted the right ascension ($\alpha$), 
declination ($\delta$) and the distance ($D$) into the corresponding Cartesian 
coordinates $(x,y,z)$.  
Let us consider the Cartesian coordinate system $(x,y,z)$ which has the origin 
in the center of the SMC at $(\alpha,\delta,D)=(\alpha_{0},\delta_{0},D_{0})$. 
The $z$ axis is pointed towards the observer. $x$-axis 
is antiparallel to the $\alpha$-axis, the $y$-axis is parallel to the 
$\delta$-axis.	$D_{0}$ is the distance between the center of the SMC and the 
observer. $D$ is the observer-source distance. $(\alpha_{0},\delta_{0})$ are 
the equitorial coordinates of the center of the SMC.  The center of the SMC is 
taken as $(\alpha_{0},\delta_{0}) = (0^{\rm h}53^{\rm m}31^{\rm s},-72^{\circ}59^{\prime}15^{\prime\prime}.7)$ \citep{smit12}. The $(x,y,z)$ coordinates are 
obtained using the transformation equations
\citep{vand101,wein01}:
\begin{eqnarray*}
\label{proj}
x=-D\sin(\alpha-\alpha_{0})\cos{\delta}, \\
y=D\sin{\delta}\cos{\delta_{0}}-D\sin{\delta_{0}}\cos{(\alpha-\alpha_{0})}\cos{\delta}, \\
z=D_{0}-D\sin{\delta}\sin{\delta_{0}}-D\cos{\delta_{0}}\cos{\alpha-\alpha_{0}}\cos{\delta}.
\end{eqnarray*}
The coordinate system of the SMC disk $(x^{\prime},y^{\prime},z^{\prime})$ is 
the same as the orthogonal system $(x,y,z)$, except that it is rotated around 
the $z$-axis by the position angle $\theta$ counterclockwise and around the 
new $x$-axis by the inclination angle $i$ clockwise. 
The coordinate transformations are \citep{vand101,wein01}:
\begin{align*}
\begin{bmatrix} x^{\prime} \\ y^{\prime} \\ z^{\prime} \end{bmatrix} 
=  \begin{bmatrix} 
\cos{\theta} & \sin{\theta} & 0  \\
-\sin{\theta}\cos{i} & \cos{\theta}\cos{i} & -\sin{i} \\
-\sin{\theta}\sin{i} & \cos{\theta}\sin{i} & \cos{i} 
\end{bmatrix}\begin{bmatrix} x\\ y\\ z \end{bmatrix}
\end{align*}
The errors $(\sigma_{x},\sigma_{y},\sigma_{x})$ in the Cartesian coordinates 
were obtained following the propagation of error formula \citep{bevi03}. 
Observed distribution of RRab stars in the SMC was modeled by a triaxial 
ellipsoid. Properties of the ellipsoid are obtained from the moment of 
inertia tensor using the principal axes transformation method \citep{deb14}.
The axes ratios $\frac{S_{i}}{S_{0}},\text{where}~i=0,1,2$, inclination of the 
longest axis along the line of sight $(i)$, position angle of line of nodes $(\theta_{lon})$ along with their associated errors were calculated using the Monte Carlo simulations. The Monte Carlo simulations for finding errors in the 
geometrical parameters were obtained with the assumption that the probability 
distribution of errors in the parameters are Gaussian and the errors in 
the obtained $(x,y,z)$ values are reliable. The observed projected Cartesian 
coordinates $(x,y,z)$ were simulated with the following steps:
\begin{enumerate}
\item [(1)] The observed Cartesian coordinates $(r_{i},\sigma_{r_{i}})$ were 
obtained from the transformations connecting them with the given 
($\alpha_{i}, \delta_{i},D_{i}$) along with the application of propagation of 
error formulae \citep{bevi03}, where $r_{i}=(x_{i},y_{i},z_{i})$ and 
$\sigma_{r_{i}}=(\sigma_{x_{i}},\sigma_{y_{i}}, \sigma_{z_{i}})$. 
$i=1,2,\dots,N$ and $N$ is the number of data points.
\item [(2)] $r_{i}$'s were taken as the centroid of the Gaussian.
\item [(3)] Box-Muller method was then used to get Gaussian distributed points 
having $\sigma_{r_{i}}$'s as errors and $r_{i}$'s as centroids. 
Essentially, this consists of the following steps \citep{bevi03}:
\begin{eqnarray}
r_{i} = r_{i}+\sigma_{r_{i}}k_{i},  \nonumber \\
\text{with}~k_{i} = \sqrt{-2.0\log{(u_{i})}}\cos{(2.0\pi v_{i}),} \nonumber
\end{eqnarray}         
where $(u_{i},v_{i})$ are pairs of uniformly distributed random numbers.   
\end{enumerate} 
Principal axis transformation method as described in \citet{deb14} was then 
applied to the simulated Cartesian coordinates of the SMC RRab stars to 
obtain viewing angles and other geometric parameters of the galaxy. The Monte 
Carlo simulation was carried for $10^{5}$ steps and corresponding values of 
the parameters were determined in each iteration. The distribution of 
various geometric parameters were then obtained after binning with a proper 
binsize. Distribution of the geometric parameters were found to be Gaussian. 
Hence, three parameter Gaussian fitting was applied to the histogram 
distributions. The $\sigma$ values of the fitted Gaussian distributions were 
taken as errors in various parameters. Viewing angle parameters 
($i,\theta_{\text{lon}}$) and geometrical parameters such as axes ratios
($S_{0}/\overline{S_{0}}, S_{1}/S_{0}, S_{2}/S_{0}$) described in 
\citet{deb14} obtained from the entire dataset of SMC RRab stars are shown in 
Figs.~\ref{fig_all} and \ref{axis}.  Their values are listed in 
Table~\ref{par_all}.     
\begin{table*}
\begin{center}
\caption{Geometric parameters of the SMC determined from the OGLE III $I$- and 
$V$-band data using absolute magnitude values determined from the Fourier 
method and the relation given by \citet{cate08}.}
\begin{tabular}{|c|c|c|c|c|c|c|c|}
\hline
&&&\multicolumn{5}{|c|}{Geometric parameters}\\
&&&\cline{1-5} 
&&Method&$S_{0}/\overline{S_{0}}$&$S_{1}/S_{0}$&$S_{2}/S_{0}$&$i[^{\circ}]$&$\theta_{\text{lon}}[^{\circ}]$ \\
\multirow{2}{22mm}{Band} &\multirow{2}{*}{I} &MVF&$1.000\pm0.004$&$1.334\pm0.007$&$08.884\pm0.123$&$1.713\pm0.128$&$74.063\pm0.510$  \\ & & MVC&$1.000\pm0.004$&$1.327\pm0.008$&$09.153\pm0.154$&$1.912\pm0.136$&$73.913\pm0.589$  \\ &\multirow{2}{*}{V} &MVF&$1.000\pm0.002$&$1.281\pm0.006$&$07.067\pm0.083$&$2.860\pm0.136$&$72.848\pm0.496$  \\& &MVC&$1.000\pm0.003$&$1.243\pm0.008$&$07.278\pm0.114$&$3.637\pm0.171$&$72.516\pm0.714$  \\
\hline
\end{tabular}
\label{par_all}
\end{center}
\end{table*}

\begin{figure*}
\vspace{0.02\linewidth}
\begin{tabular}{cc}
\vspace{+0.01\linewidth}
  \resizebox{0.46\linewidth}{!}{\includegraphics*{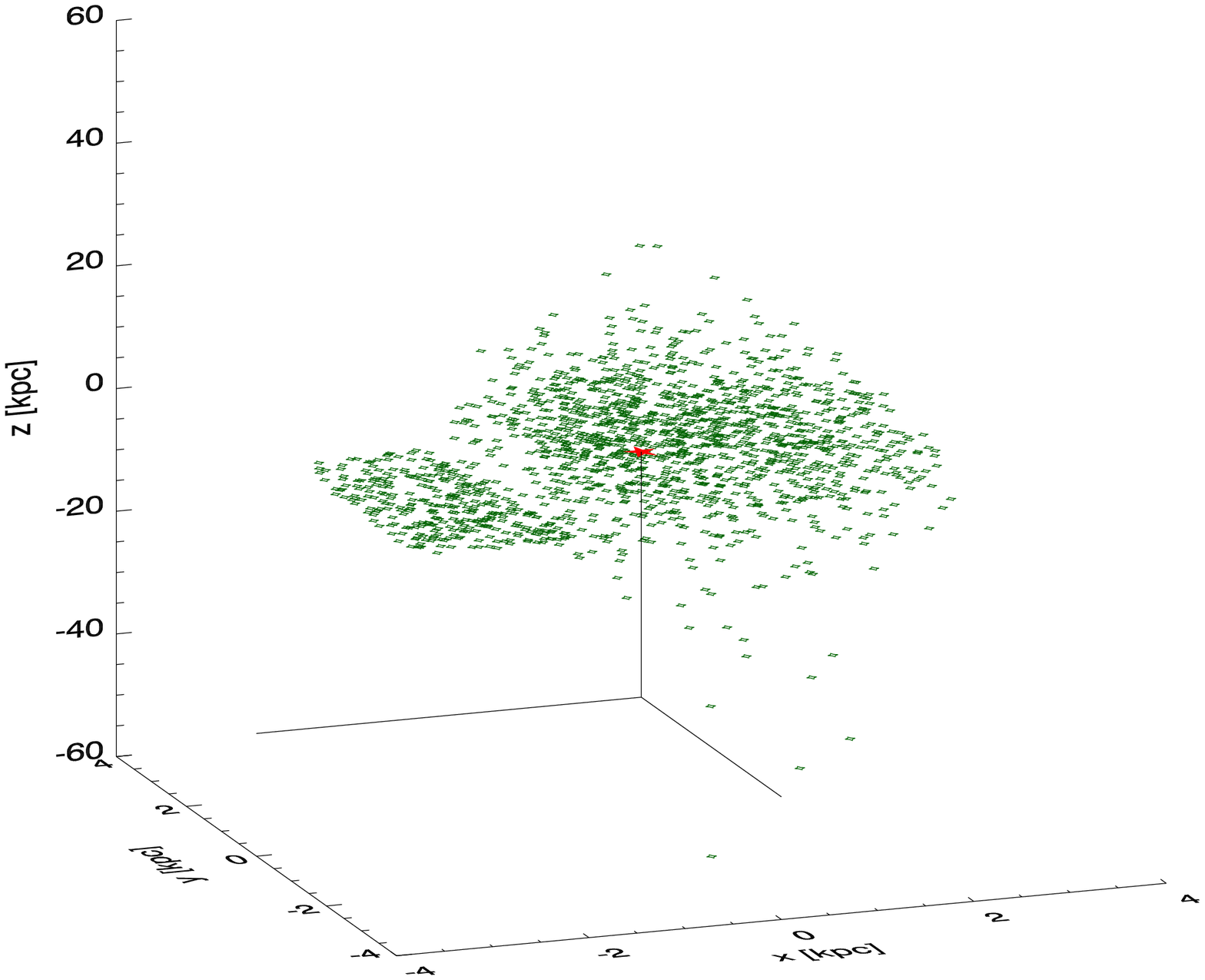}}&
  \resizebox{0.46\linewidth}{!}{\includegraphics*{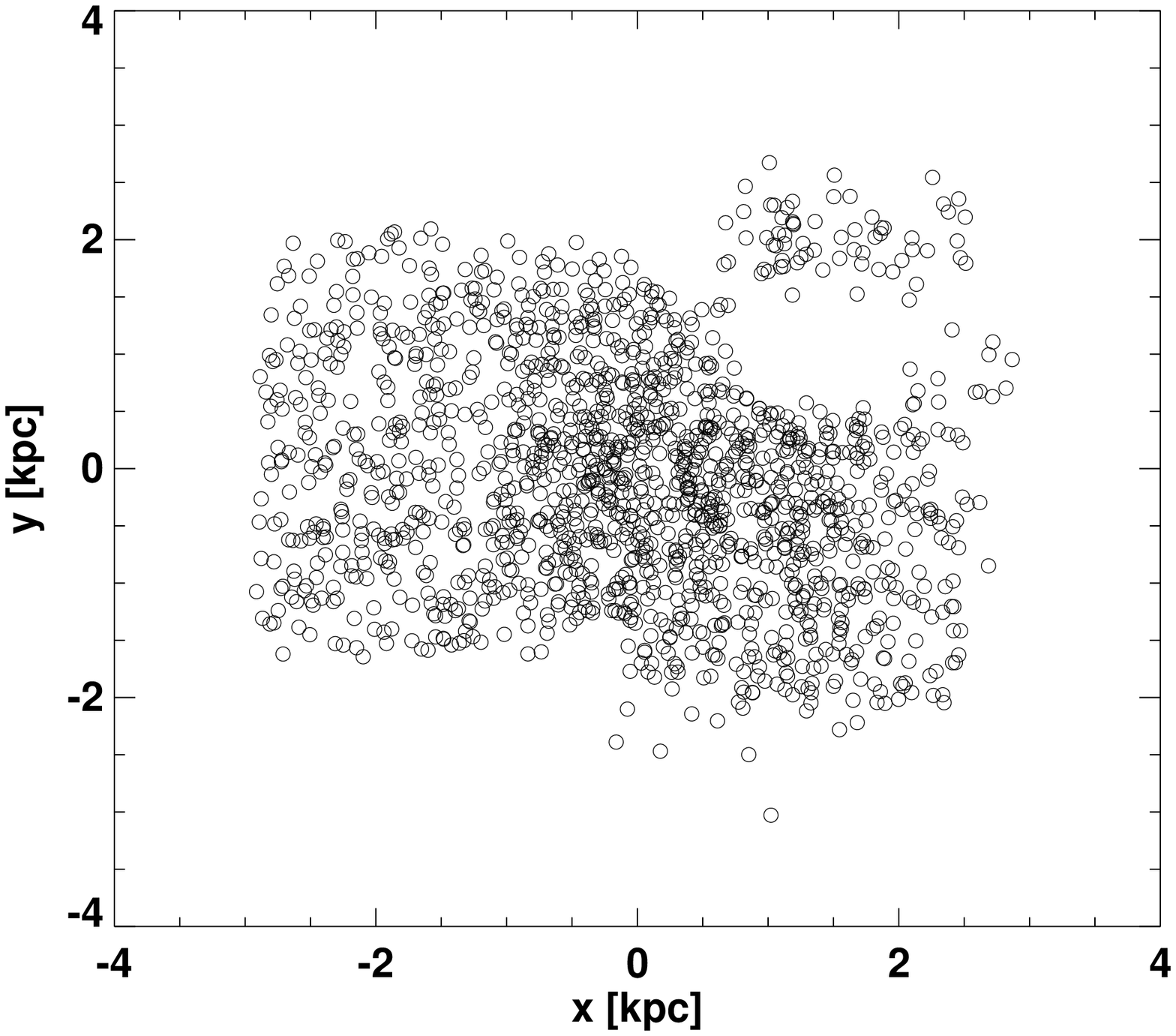}}\\
\vspace{0.01\linewidth}
\resizebox{0.46\linewidth}{!}{\includegraphics*{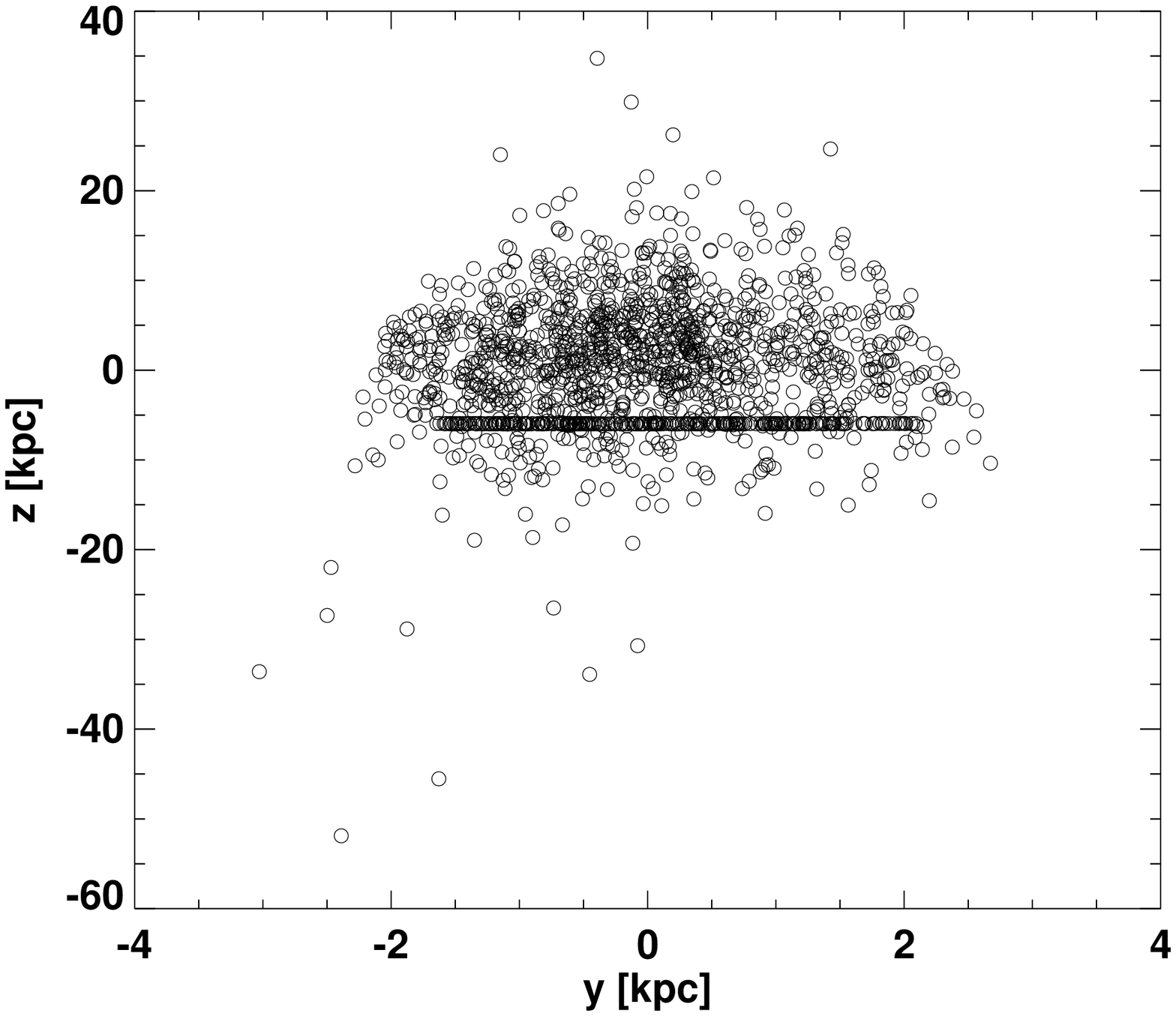}}&
\resizebox{0.46\linewidth}{!}{\includegraphics*{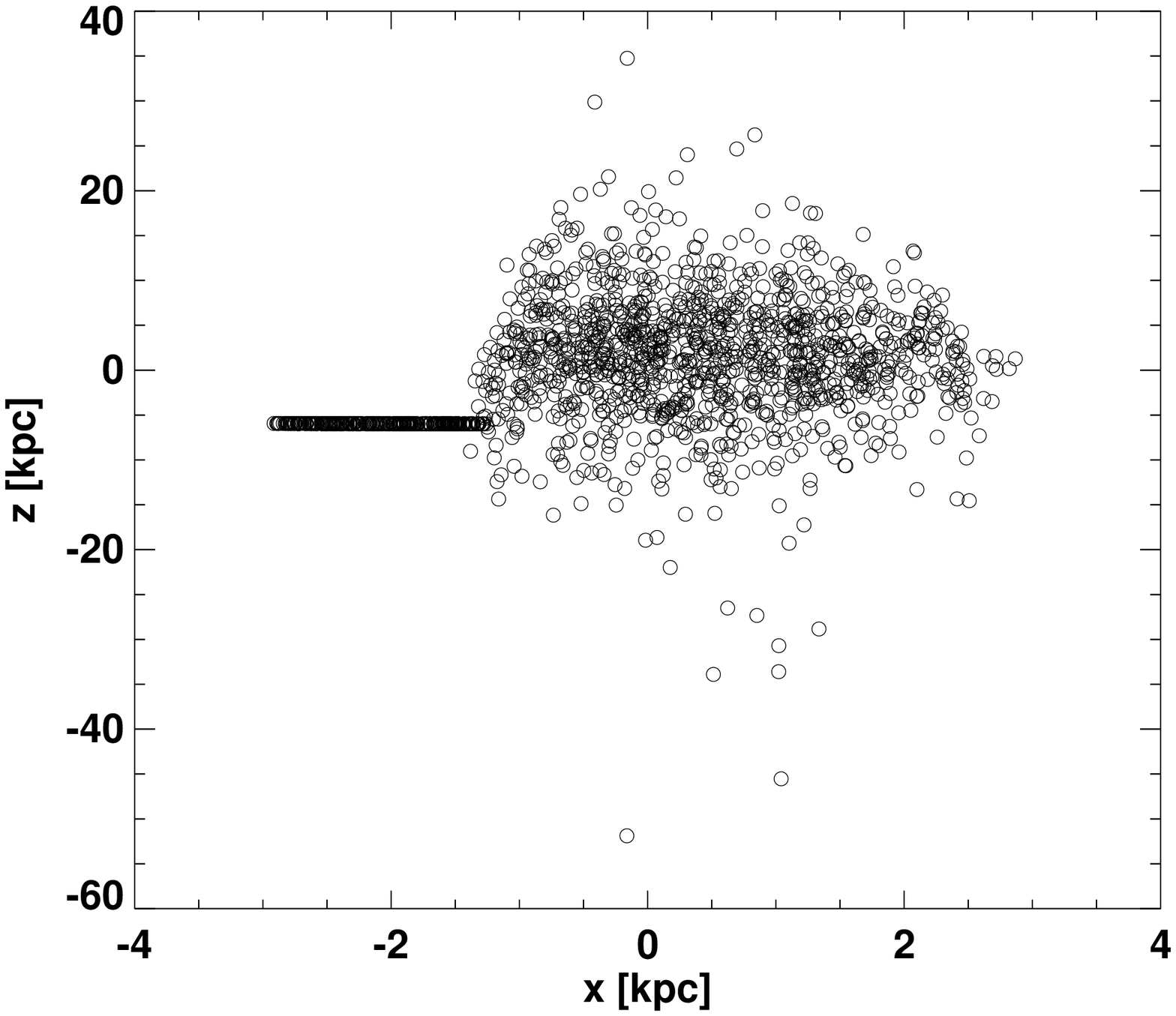}}\\
\vspace{-0.04\linewidth}
\end{tabular}
\caption{Upper left panel shows separation of SMC into its northeast arm 
(shown as planar distribution) and that of the main body (shown as ellipsoidal 
distribution). The centre of the galaxy is marked by red star symbol. Other 
panels depict the projected distributions on the XY, YZ and XZ planes as 
well. The NE arm appears as a line in the two-dimensional YZ and XZ 
projections.}  
\label{separation}
\end{figure*}

\begin{figure*}
\vspace{0.02\linewidth}
\begin{tabular}{cc}
\vspace{+0.01\linewidth}
  \resizebox{0.46\linewidth}{!}{\includegraphics*{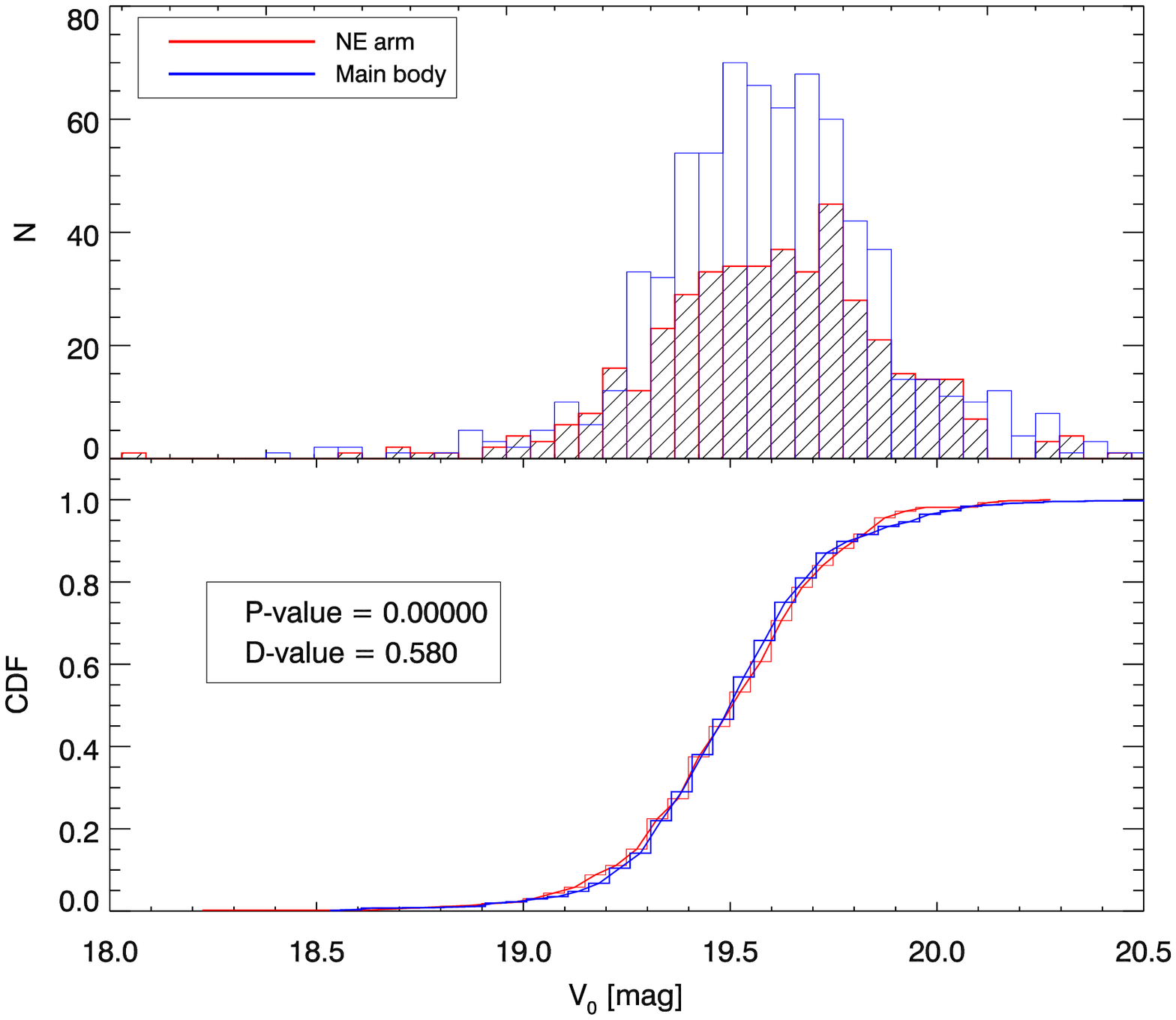}}&
  \resizebox{0.46\linewidth}{!}{\includegraphics*{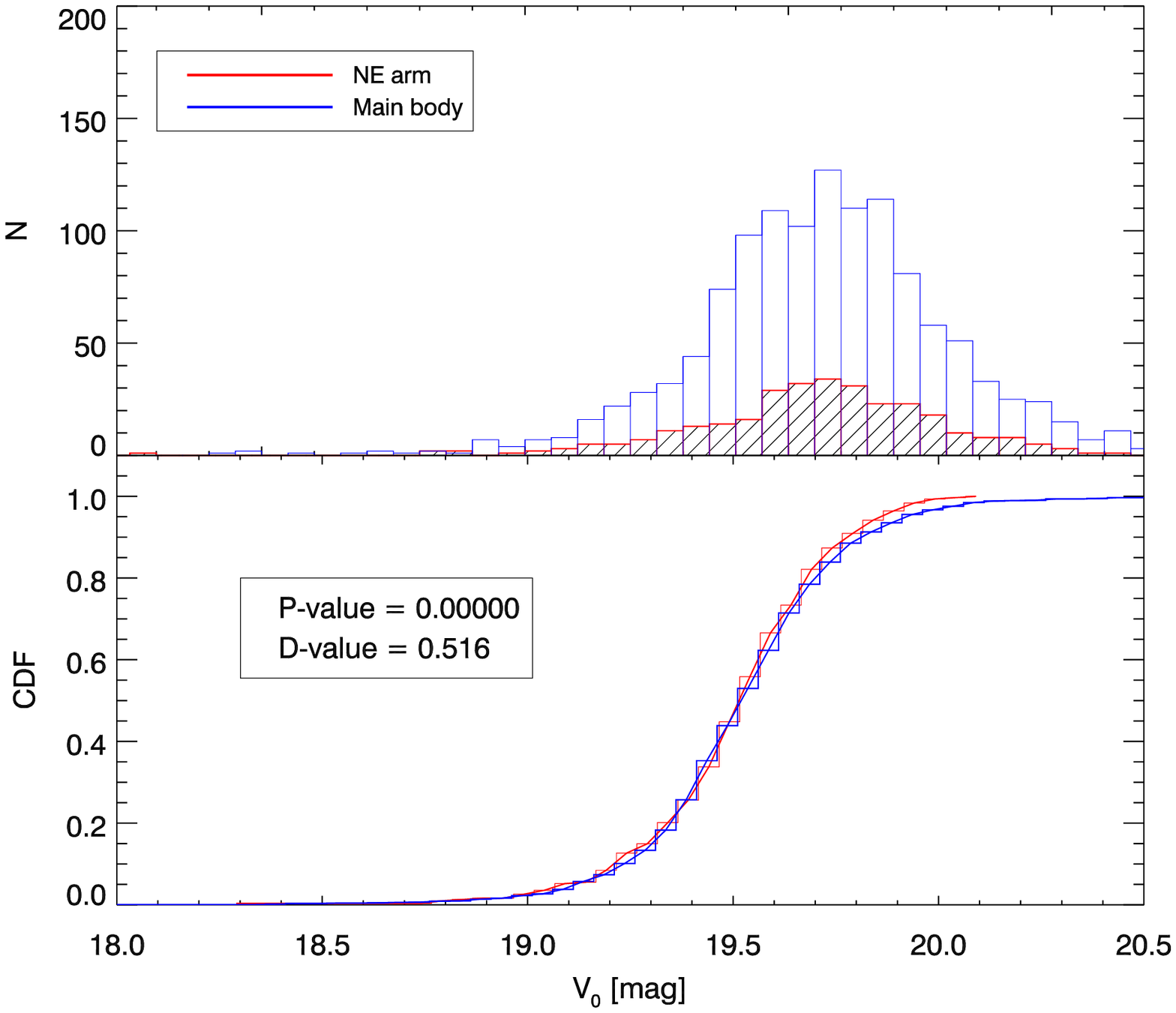}}\\
\vspace{+0.01\linewidth}
  \resizebox{0.46\linewidth}{!}{\includegraphics*{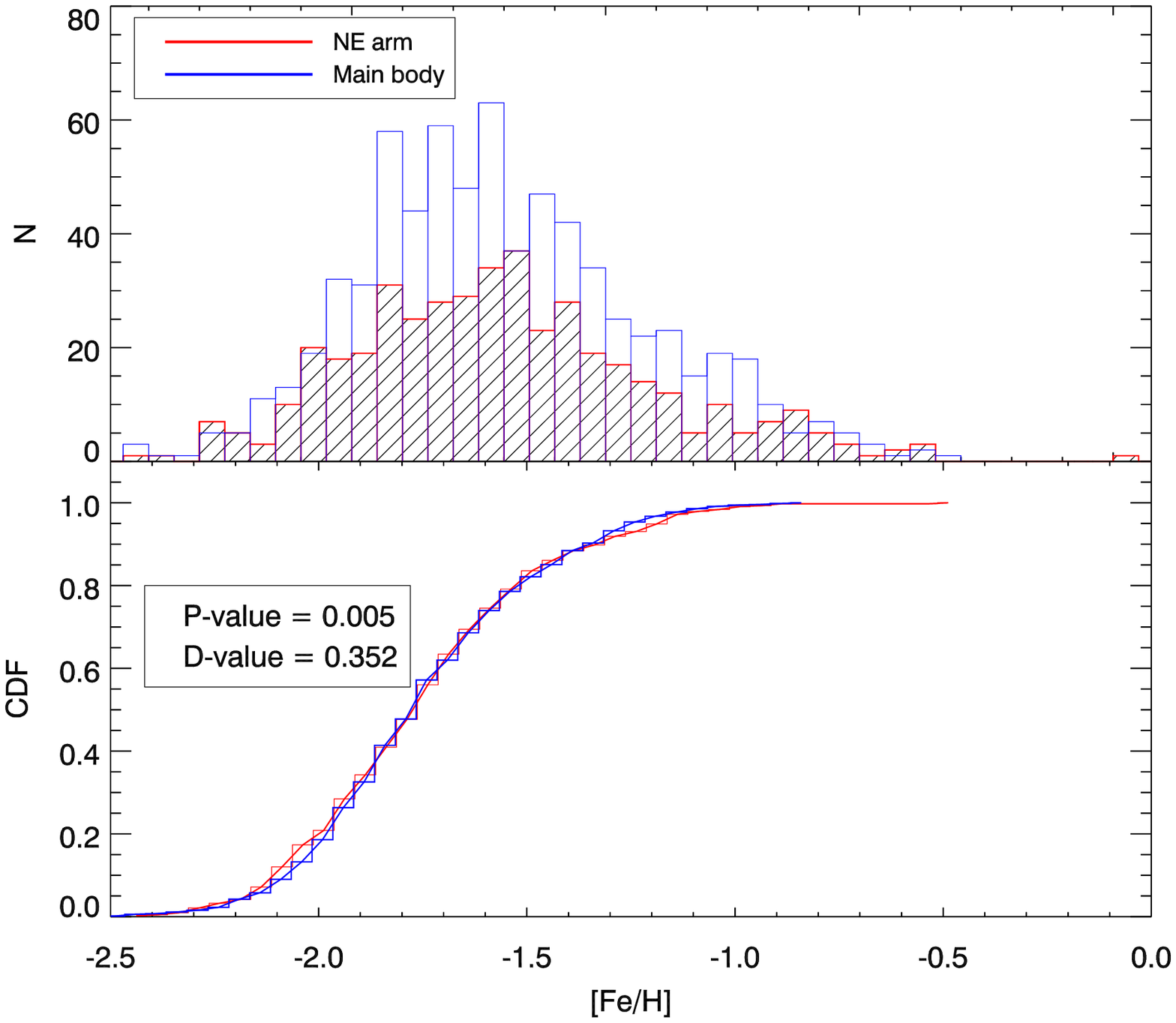}}&
  \resizebox{0.46\linewidth}{!}{\includegraphics*{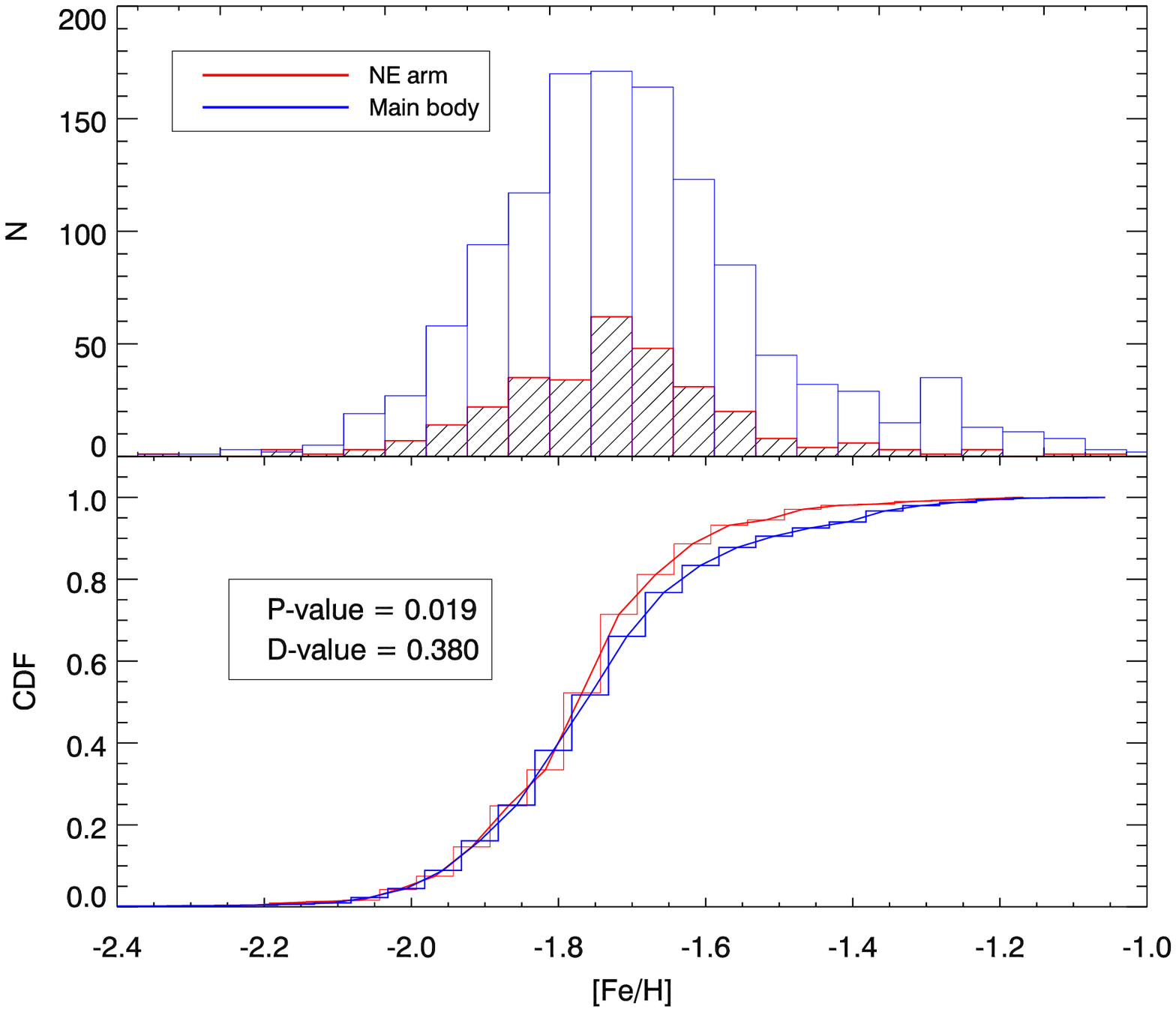}}\\
\vspace{-0.04\linewidth}
\end{tabular}
\caption{Cumulative distribution of extinction corrected RRab magnitudes 
$V_{0}$  and metallicities $[Fe/H]$ for the RRab stars (left panel shows the 
use of $V$-band directly on the empirical relation, whereas, right panel 
indicates the use of $I$-band data calibrated to the $V$-band). NE arm and  
SMC main bosy are marked with red and blue color lines, 
respectively.}
\label{cumulate}
\end{figure*}

From the Cartesian three-dimensional distribution of RRab stars as shown in 
Fig.~\ref{separation}, it is quite evident that north east (NE) arm is nearer 
to us than the SMC main body. In order to find exact location 
of NE arm of the SMC, we have used an R\footnote{\url{http://www.r-project.org/}} statistical package 
{\it{changepoint}} for changepoint 
analysis \citep{rebe1,rebe2}. The function cpt.meanvar() was used to find 
changes in the mean and variance for $z$ distribution. Locations of main body 
and NE arm of the SMC were thus identified from the changepoint analysis. 
Principal axis transformation method was then applied to the SMC main body 
since its distribution resembles a three-dimensional ellipsoid. This is done 
in four cases (Two in each bands corresponding to the methods `MVF' 
and `MVC'). Results obtained from the principal axis transformation method are 
shown in Table~\ref{main-bar}. On the other hand, NE arm of the SMC resembles 
like a plane. Therefore, plane-fitting procedure was applied to find its 
viewing angle parameters such as inclination and position angle of line of 
nodes. NE arm of the SMC was fitted with a plane-fitting equation of 
the form \citep{niko04}
\begin{equation}
\label{planefit}
z_{i}=c+ax_{i}+by_{i},~i=1,2,\dots,N,
\end{equation}                                         
where $N$ is the number of data points. 
The inclination angle ($i[^{\circ}]$) can be obtained from 
\begin{displaymath}                         
i=\arccos{\left(\frac{1}{\sqrt{(1+a^2+b^2)}}\right)},
\end{displaymath}
and if we define $\gamma = \arctan(|a|/|b|)$, then
\begin{align*}
\theta_{\text{lon}} &=
\begin{cases}
\gamma & \text{if}~~a < 0~~\text{and}~~b > 0, \\
\gamma & \text{if}~~a > 0~~\text{and}~~b < 0,\\
\frac{\upi}{2} + \gamma & \text{if}~~a < 0~~\text{and}~~ b < 0, \\
\frac{\upi}{2} + \gamma & \text{if}~~a > 0~~\text{and}~~ b > 0, \\
0 & \text{if}~~a = 0~~\text{and}~~b~~\neq 0, \\
\text{sign}(a)\frac{\upi}{2} &~~\text{if}~~a \neq 0~~\text{and}~~b = 0,\\
\text{undef.} &~~\text{if}~~a = b = 0.
\end{cases}
\end{align*}
A weighted plane-fitting procedure using mpfitfun in IDL was used to fit NE 
arm of the SMC \citep{mark09,mark12}. Results of the plane-fitting procedure 
have been listed in Table~\ref{main-bar}. Errors in $i$ and 
$\theta_{\text{lon}}$ were obtained from the propagation of error formula.
From Table~\ref{main-bar}, one can find that the values of $i$ are comparable 
to each other within the error bars 
indicating that the values obtained using two different datasets and empirical 
relations are reliable. On the other hand, if we compare the values of viewing 
angle parameters, then we find that $\theta_{\text{lon}}$ values of the SMC 
main body are significantly different for the two bands. Difference in the 
$\theta_{\text{lon}}$ arises due to poor sampling in distance distribution in 
the case of $V$-band as compared to those in the $I$-band. We take the 
results obtained from the $I$-band data to be more reliable because of larger
number of stars and better phase coverage of light curves as compared to 
the corresponding $V$-band light curves. The viewing angle parameters obtained 
for NE arm of the SMC indicate that it is slightly misaligned with the SMC 
main body.                
\begin{table*}
\begin{center}
\caption{Geometric parameters of the SMC determined from the OGLE III $I$- and 
$V$-band data using absolute magnitude values determined from the Fourier 
method and the relation given by \citet{cate08}.}
\label{main-bar}
\begin{tabular}{|c|c|c|c|c|c|c|c|}
\hline
&&&\multicolumn{5}{|c|}{Geometric parameters}\\
&&&\cline{1-5} 
&&Method&$S_{0}/\overline{S_{0}}$&$S_{1}/S_{0}$&$S_{2}/S_{0}$&$i[^{\circ}]$&$\theta_{\text{lon}}[^{\circ}]$ \\
\multirow{2}{22mm}{SMC Main Body (Ellipsoid-fit)} &\multirow{2}{*}{I} &MVF&$1.000\pm0.003$&$1.185\pm0.004$&$09.764\pm0.144$&$0.753\pm0.093$&$55.569\pm0.771$  \\ & & MVC&$1.000\pm0.004$&$1.186\pm0.005$&$09.916\pm0.188$&$0.573\pm0.106$&$55.552\pm0.949$  \\ &\multirow{2}{*}{V} &MVF&$1.000\pm0.004$&$1.346\pm0.007$&$10.221\pm0.149$&$0.801\pm0.106$&$82.799\pm0.401$  \\& &MVC&$1.000\pm0.006$&$1.342\pm0.009$&$10.839\pm0.254$&$0.354\pm0.106$&$83.068\pm0.551$  \\ \hline
\multirow{2}{22mm}{SMC NE Arm (Plane-fit)} &\multirow{2}{*}{I} &MVF&$-$&$-$&$-$&$2.254\pm0.856$&$85.937\pm14.030$  \\ & & MVC&$-$&$-$&$-$&$2.254\pm0.864$&$85.932\pm14.149$  \\ &\multirow{2}{*}{V} &MVF&$-$&$-$&$-$&$1.284\pm0.210$&$83.710\pm09.309$  \\& &MVC& $-$ & $-$ & $-$ &$0.500\pm0.193$&$78.744\pm22.445$ \\
\hline
\end{tabular}
\end{center}
\end{table*}
Cumulative distribution of extinction corrected magnitudes $(V_{0})$  and 
metallicities ($[Fe/H]$) of NE arm (sample 1) and SMC main body (sample 2) are 
shown in Fig.~\ref{cumulate}. In order to test whether the two 
samples came from the same distribution at a significance level of $0.05$, we 
use two sample Kolomogorov-Smirnov (KS) statistics \citep{pres02}. The 
KS statistics looks at the maximum absolute difference between the empirical 
CDF of sample 1 and empirical CDF of sample 2. It is a nonparametric test that 
allows statistical comparison of two one-dimensional distributions. 
The KS test applied to the two samples yields $D$-value and $P$-value.  
Here, $D$ is the maximum value of the absolute difference between the two 
CDFs.  From Fig.~\ref{cumulate}, one can find that since the $P$-value is 
greater than our assumed significance level of $0.05$, we reject the null 
hypothesis and conclude that the cumulative distribution of sample 1 is 
significantly different from sample 2, with sample 1 being brighter and 
metal rich as compared to sample 2. Application of planefit procedure to 
the $V$-band data directly with distance determined using the MVF and MVC 
methods yields the value of $c$ in Eqn.~\ref{planefit} to be $-5.949\pm0.233$ 
kpc and $-8.087\pm0.364$ kpc, respectively. 
Corresponding values of $c$ obtained from the  $I$-band data calibrated to the
$V$-band are $-5.299\pm0.509$ kpc and $-6.020\pm0.847$ kpc. Negative values of 
$c$ obtained in each of the four cases indicate that NE arm of the SMC is
nearer to us than the plane of the SMC main body. Since in majority of the 
cases, we get the NE arm to be nearer to us by $\sim 6$ kpc, we take this as 
the tentative estimate. It has also been found from the analyses that 
residuals of the planefit reveal a highly symmetric hyperbolic paraboloid warp 
of low amplitude ($\sim 0.03$ kpc). In order to test whether the results 
obtained so far are an artifact or a real feature of the SMC, we resort to the 
metallicity relation of \citet{smol05} using $I$-band data directly  and 
absolute magnitude relation of \citet{cate08} as followed in 
section~\ref{smol}. Indeed, it has been found that same patterns are reflected 
and similar results are obtained in the resulting analyses.        
\section{line of sight depth of the smc}
\label{los}
Line of sight (LOS) of the SMC has been studied using numerous tracers in 
the literature  and is an important parameter to ascertain the extent of the 
SMC \citep{crow01,subr09,smit12,kapa12,hasc_smc}. RR Lyrae stars in the 
present analysis have been used as proxies to get an estimate of the LOS 
depth. When combined with the metallicity values, 
it serves as a robust parameter to the underlying different galactic 
substructures \citep{crow01,kapa12}. To estimate the  intrinsic line of sight 
depth of the SMC, we take into account the standard deviation 
$(\sigma_{\rm obs})$ distribution of the RRab distances and measurement errors 
($\sigma_{\rm error}$) in the individual distances. The overall standard 
deviation $(\sigma_{\rm obs})$ is a combination of these two factors in 
quadrature \citep{crow01}:
\begin{equation}
\sigma^{2}_{\rm obs} =\sigma^{2}_{\rm int}+\sigma^{2}_{\rm error}.  
\end{equation}                    
The intrinsic $\pm 1\sigma$ line of sight depth is then given by 
$\left(\sigma^{2}_{\rm obs} - \sigma^{2}_{\rm error}\right)^{2}$.  
Using the $I$-band data, we have found the line of sight depth to be $1\sigma_{los} = 4.91\pm0.65, 4.12\pm0.71$ kpc from the distances determined 
using MVF and MVC, respectively. Corresponding values of these 
parameters in the $V$-band were found to be $5.34\pm0.61$ and $4.63\pm0.83$ 
kpc, respectively. From a study of $454$ RRab stars in the $V$-band, 
\citet{kapa12} determined the LOS depth to be $\sigma_{int} = 5.3\pm0.4$ kpc. 
It may be noted that the LOS depth estimate has a lower value when 
\citet{cate08} relation is used to derive the line of sight distances then when 
\citep{kova01} relation is used. The variance weighted mean for the above 
estimates $(\text{weights} = \frac{1}{\sigma_{2}})$ yields 
$\sigma_{int} = 4.81\pm 0.53$ kpc \citep{bevi03}. There is a caveat in the 
LOS depth estimation from the optical band photometric light curve data as it 
is highly sensitive to the interstellar reddenings. A large sample of near 
infrared (NIR) photometric light curve data can help us ascertaining the exact 
LOS depth of the SMC in different directions and thus help in locating the 
various components of the SMC. 
Use of different reddening maps applied to the optical band light curve data 
also yields different estimates of the LOS depth values \citep{crow01}. 
However, most of the estimates of the $1\sigma$ LOS depth values are limited 
in the range between $4$ and $6$ kpc \citep{crow01,smit12,hasc_smc,kapa12}. 
Despite completely different ways of obtaining the estimate of the LOS depth 
of the SMC from two OGLE bands, its value is consistent and  well within the 
range. Combined with the LOS depth and the  metallicity values of the sample, 
\citet{kapa12} found the existence of different structures consisting of 
different populations. Dividing the sample on the basis of their 
metallicities, they found that the metal rich stars show small variations in 
their LOS depth while the metal poor stars exhibit larger 
LOS depth variations. From their analysis, they also concluded that the inner
regions of the SMC consists of a thicker structure mimicking a bulge. Metal 
rich and metal poor stars seem to belong to different dynamical structures. 
Metal poor stars have a smaller scale height and may belong to a thick disk and 
metal poor stars seem to belong to the halo. However, the results obtained by 
\citet{kapa12} should be taken with caution since these have been obtained 
from a smaller sample of $454$ RRab stars with their $V$-band light curves 
and may suffer from selection effect.  In order to test the results of 
\citet{kapa12}, we have also investigated the LOS depth of the sample using 
the $I$-band data by dividing the sample into the metal poor 
($[Fe/H] < \overline{[Fe/H]}-\sigma =-1.85$ dex) and metal rich stars 
($[Fe/H] > \overline{[Fe/H]}+\sigma = -1.60$ dex). 
In order to find if there is any variation in the LOS depth of the two samples 
belonging to the different metallicity groups at a significance level of 
$0.05$, we again perform a two sample KS test using the $I$-band data 
calibrated to the $V$-band. The two sample KS test yields $D = 0.110$ and 
$P$-value = $0.90$. The $P$-value $> 0.05$ indicates that we fail 
to reject the null hypothesis implying that there is no evidence in 
the data to suggest that the two CDFs are different. This is contrary to that 
obtained by \citet{kapa12}. While calculating the LOS depth, it has been found 
that observed uncertainties in the distance measurement 
($\sigma_{\text{err}}$) of a few stars are greater than the observed depth 
($\sigma_{\text{obs}}$). The LOS depth determinations for those stars have 
been left out. 

To investigate the LOS depth variations in the SMC, we have divided 
the sample into three parts, viz., NE ($x<-0.5,~y>0.5$ kpc), central 
($-0.5<x<0.5,-0.5<y<0.5~\text{kpc})$ and SW ($x > 0.5~,y<-0.5$~\text{kpc}). 
In order to see the variation of depth among the three different 
regions of the SMC, we study their normalized LOS CDF distribution through KS 
test. A two sample KS test indicates that that LOS depth of the NE and SW 
parts of the SMC are not significantly different. On the other hand, KS test 
applied on LOS CDF distribution of the central part and other parts 
combined (NE+SW) yields $D=0.316$ and $P$-value=$0.008$. These values indicate 
that the difference in CDF of the two distributions are different at a 
significance level of $0.05$, with LOS depth of central part being larger than 
rest of the parts of the SMC. This indicates that SMC may have a bulge. 
\section{Metallicity gradient}
\label{metgrad}
The presence of radial metallicity gradient in a galaxy indicates the  
presence of different stellar populations which are responsible for 
formation and evolution of the galaxy \citep{bern08,cion09}. The detection 
of metallicity gradient in the SMC using the CaII triplet (CaT) spectroscopy 
was first reported by \citet{carr08} using the $350$ red giant branch stars of 
$13$ SMC fields distributed in the range $1^{\circ}-4^{\circ}$ from the SMC 
center, wherein the metallicity decreases from the center of the galaxy 
towards the outermost regions. In their search for the metallicity gradient they also 
found that this gradient is related to an age gradient where the stars 
concentrated in the central regions are generally younger and metal rich.  
Recent studies attempt to bear out the presence of this gradient. Using the 
CaT metallicity study of the SMC field of $3037$ red giant stars spread across 
$37.5$ deg$^{2}$ centred on the galaxy, \cite{dobb14} found evidence of 
metallicity gradient of $-0.075\pm0.011$ dex/deg over the inner $5$ deg 
suggesting an increasing number of young stars with decreasing galacto-centric 
radius.  However, the formal errors in their determinations of $[Fe/H]$  are 
too optimistic. Finding meagre gradients as in the case of the SMC using the 
$[Fe/H]$ values requires careful estimations of their errors and determining 
the weighted mean metallicities and their errors while binning the 
galacto-centric radius. The weighted mean metallicities and the errors in each 
bin tells how statistically significant the gradient is. Using a sample of 
$454$ RRab stars in $V$-band from the OGLE-III database, \citet{kapa12} found  
a metallicity gradient of $-0.013\pm0.007$ dex/kpc with increasing metal 
abundance towards the dynamical center of the SMC. However, the contrary was 
found by \citet{cion09} using the $[Fe/H]$ derived from the ratio of C- and 
M-type AGB stars analysed as a function of galacto-centric distance. A 
constant value of $[Fe/H] =-1.25\pm0.01$ dex up to $12$ kpc was obtained by 
\citet{cion09}. The result of no metallicity gradient was reinforced by the 
study of the $25$ SMC clusters on a homogeneous metallicity scale and with 
relative small metallicity errors by \citet{pari09}. Different values of 
metallcity gradients using various tracers obtained in the literature may be 
related to their redistributions from the positions they were formed 
\citep{rosk08,cion09}. Comparison of metallcities of SMC field and cluster 
stars similarly may also yield significant metallcity gradients due to the 
fact that the field stars are more metal poor than ther corresponding cluster 
stars surrounding them \citep{pari09}.          
 
\begin{figure}
\includegraphics[width=0.46\textwidth,keepaspectratio]{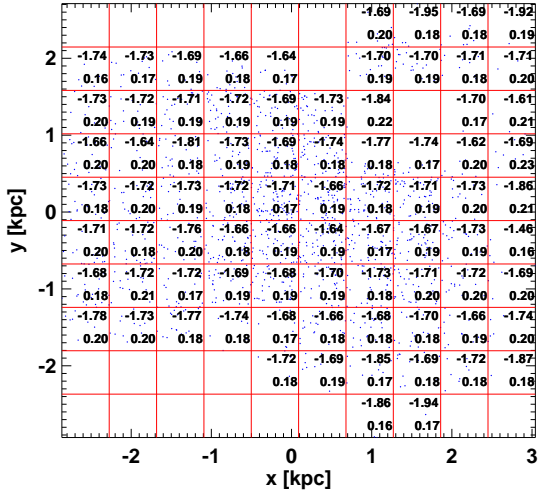}
\caption{Metallicity distribution of RRab stars in the SMC.  Metallicity 
values are binned on a $10\times 10$ coordinate grid. In each bin, average 
metallicities and their associated errors are calculated which are due to 
uncertainties in the determination of Fourier parameter $\phi_{31}$ and are 
shown in each grid box.}
\label{metmap}
\end{figure}

\begin{figure}
\includegraphics[width=0.46\textwidth,keepaspectratio]{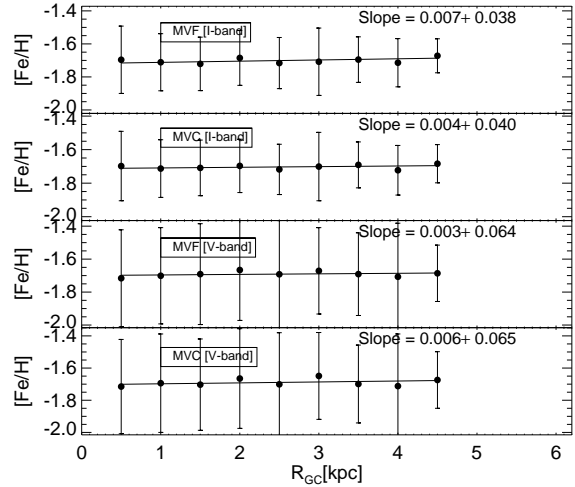}
\caption{Mean metallicity distribution of RRab stars as a function of the 
galacto-centric distance $(\rm R_{\rm GC})$ in kpc with a bin size of $0.5$ 
kpc. Mean metallicities in each bin have been obtained using the \citep{jurc96} 
empirical relations in the $V$ and $I$-band with distances measured using the 
two methods described in the text.}
\label{metgrad_fig}
\end{figure}

In the investigation of any gradient in a galaxy, it is important to determine 
viewing angles of the galaxy accurately. In this paper, we have 
determined the geometry of the SMC using the more precise value obtained from 
the principal axis transformation applied on the de-projected Cartesian coordinates.  In order to investigate for any metallicity gradient in the SMC, the true 
galacto-centric distances were obtained using the procedure described by 
\citet{cion09}. The coordinate system was rotated  according to 
\begin{align}
X^{\prime} = X\cos({\theta})+Y\sin({\theta}) \\
Y^{\prime}=-X\sin({\theta})+Y\cos({\theta}).
\end{align}
The $({X^{\prime},~Y^{\prime}})$ system was then de-projected using
\begin{align}
Y^{\prime\prime} = Y^{\prime}/\cos{(i)}
\end{align} 
The angular distances were then calculated and converted into kpc with
\begin{align}
d_{\rm deg} = \sqrt{{X^{\prime}}^2+{Y^{\prime\prime}}^2}, \\
d_{\rm kpc} = D\times\tan{(d)},
\end{align}
where $d$ is the angular distance to each star. Here the Cartesian coordinates 
$(X(\alpha,\delta),Y(\alpha,\delta))$ are calculated following the method 
described in \citet{vand101}. This transformation helps
in projecting a sphere onto a plane. Metallicity distribution 
of RRab stars in the SMC is shown in Fig.~\ref{metmap}. Metallicity values are 
binned on a $10\times 10$ coordinate grid. In each bin, the average 
metallicities and the associated errors which are due to the uncertainties in 
the determination of the Fourier parameter $\phi_{31}$ are calculated. 
Fig.~\ref{metgrad_fig} depicts the variance weighted mean metallicity 
values computed for four sets as a function of galacto-centric distance in bins 
of $0.5$ kpc from the SMC center. A least-squares fit to the distance and mean 
metallicity values with their estimated errors to each of the data sets yields 
slopes of $0.007\pm0.038, 0.004\pm 0.040$, $0.003\pm0.064$ and $0.006\pm0.065$ 
dex kpc$^{-1}$.  All these values correspond to no statistically significant 
metallicity gradient. In the estimation of mean metallicity values in each 
distance bin, it has been ensured that the number of stars is greater than
$10$ for reliable statistics. Also, in the calculation of mean metallicity
errors, the errors due to the uncertainties in the Fourier parameters and the 
systematic errors in each of the empirical relations have been taken into 
account. These two errors were added quadratically for each star 
in order to estimate the mean metallicity and its associated error in a 
distance bin. All of the empirical relations for metallicity calculations do 
not show any significant metallicity gradient within the uncertainties, 
consistent with the results obtained by \citet{cion09} and \citet{pari09}.      
\section{SMC structure determination using Smolec's (2005) metallicity relation}
\label{smol}
\begin{figure}
\includegraphics[width=0.46\textwidth,keepaspectratio]{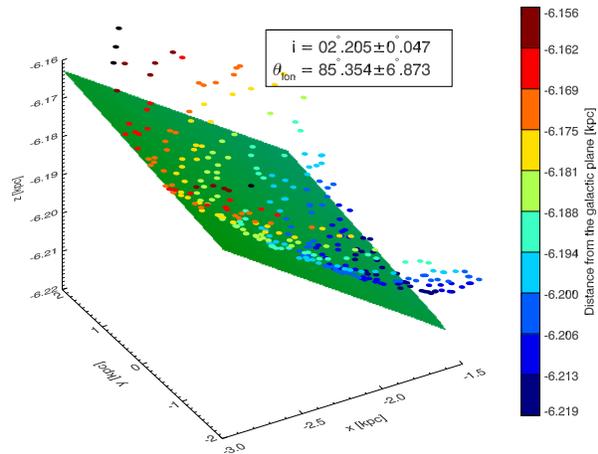}
\caption{NE part of the SMC. Solid surface in green color denotes 
planefit to the data.}
\label{nepart}
\end{figure}
\begin{figure}
\includegraphics[width=0.46\textwidth,keepaspectratio]{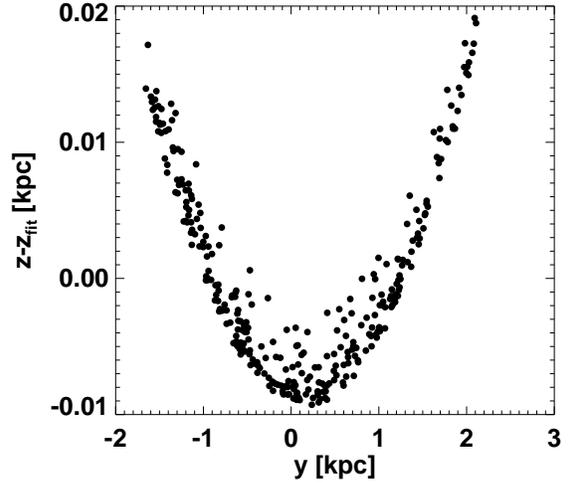}
\caption{Distribution of residuals $(z-z_{\text{fit}})$ of the NE arm 
RR Lyrae stars as a function of $y$ coordinates, where $z_{\text{fit}}$ are the 
fitted $z$-values obtained from Eqn.~(\ref{planefit}) . The distribution of 
stars take a hyperbolic paraboloid shape in $3$d because of the gravitational interation 
exerted on it by the LMC and the main body of SMC. The scatter in the $2$d 
plot is due to the projection from $3$d.}
\label{smc_residuals}
\end{figure}
\begin{figure*}
\vspace{0.02\linewidth}
\begin{tabular}{cc}
\vspace{+0.01\linewidth}
  \resizebox{0.46\linewidth}{!}{\includegraphics*{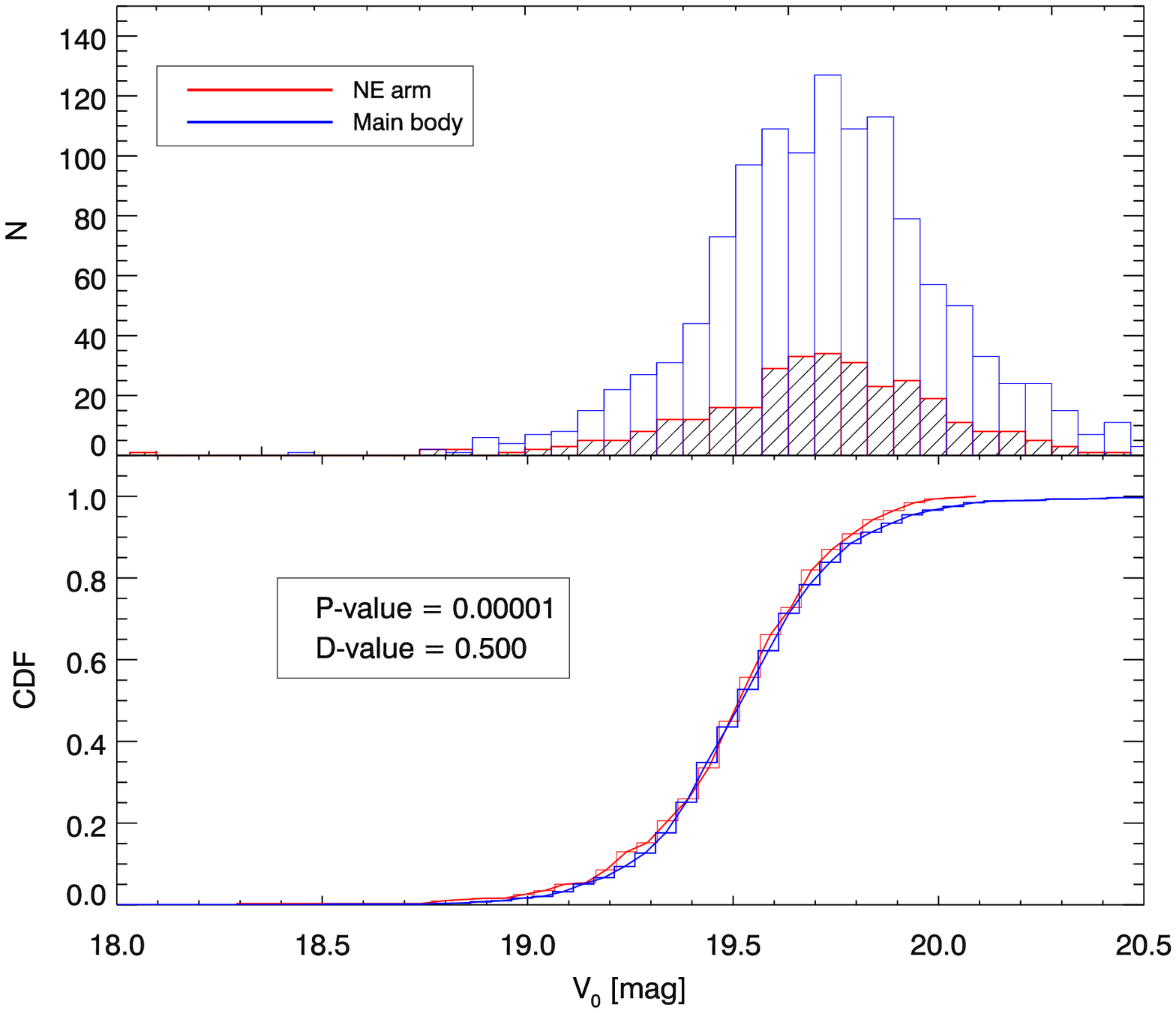}}&
  \resizebox{0.46\linewidth}{!}{\includegraphics*{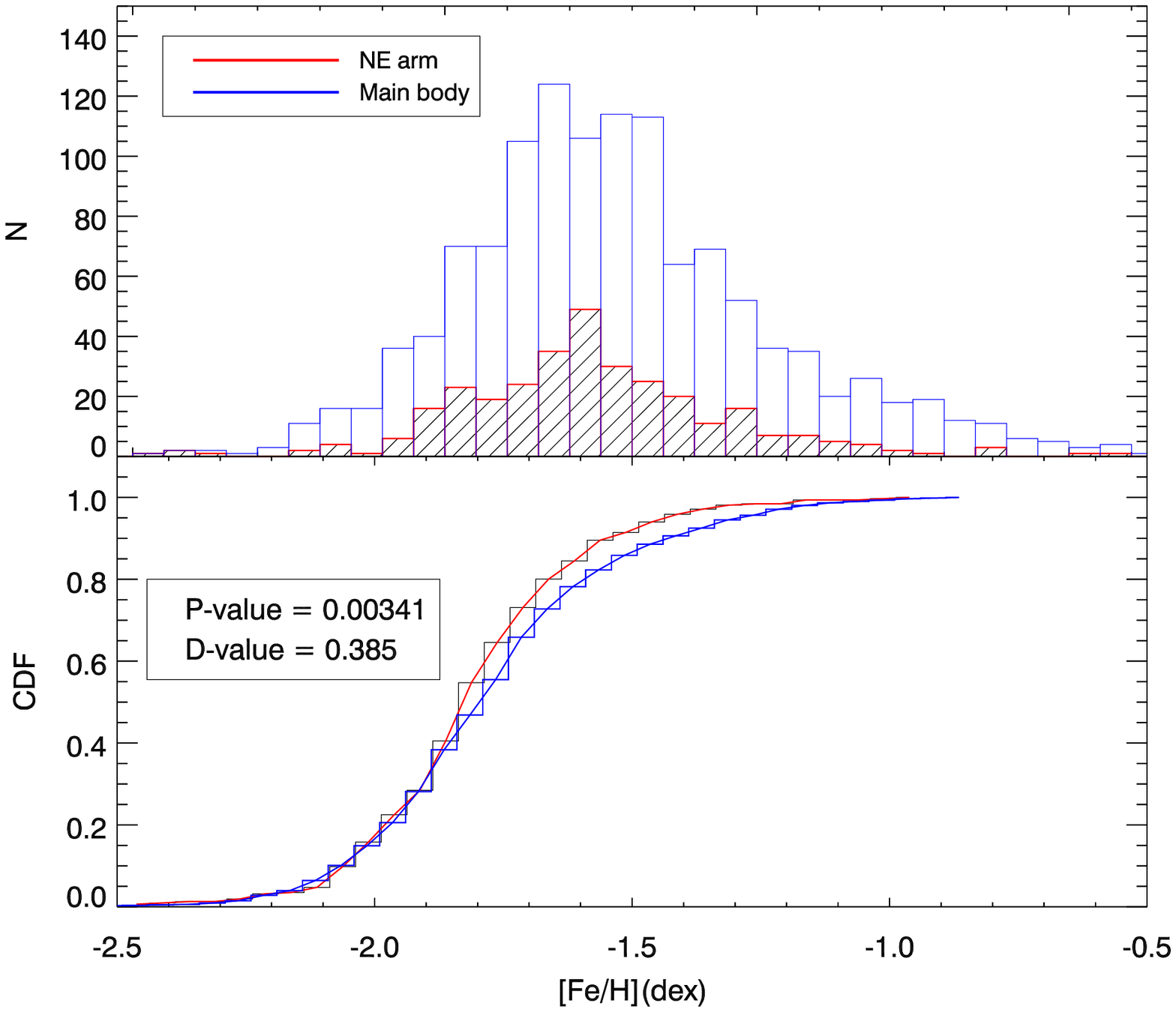}}\\
\vspace{-0.04\linewidth}
\end{tabular}
\caption{Cumulative distribution of extinction corrected RRab magnitudes 
$V_{0}$ for the RRab stars in the $V$-band with metallicities as determined 
from the \citep{smol05} relation. Stars in the northeastern arm and the SMC 
main body are marked with red and blue color lines, respectively.}
\label{hist_smol}
\end{figure*}

In order to compare, contrast and substantiate the viewing angle and 
geometrical structural parameter determinations of the SMC using $V$- and 
$I$-band data as discussed above, we now use the $M_{V}-[Fe/H]$ relation of 
\citet{cate08} with $[Fe/H]$ determined from the $I$-band light curve data 
directly using the \citet{smol05} relation. Absolute magnitude determined 
using the \citet{smol05} relation is referred to as `MVS'. 
The inclination angle and position 
angle of line of nodes for the entire sample of RRab stars obtained from the
ellipsoid fitting were found to be $i = 3^{\circ}.144\pm0^{\circ}.129$ and 
$\theta_{\text{lon}} = 74^{\circ}.856\pm0^{\circ}.514$. The axes ratios were
obtained as: $1.000\pm 0.002:1.284\pm0.006:7.474\pm0.101$. The sample of RRab 
stars are divided belonging to the SMC main body and NE arm on the basis of the 
changepoint analysis. Bimodal distribution of $z$-values also discriminates 
the NE arm from the main body of the SMC. A planefit procedure is applied to 
the NE arm which yields the value of $c =-6.251\pm0.854$ kpc. The fitted plane
is shown in Fig.~\ref{nepart}. The constant $c$ represents the shift in the 
positive or negative $z$-direction from the 
$xy$-plane of the SMC. The value of $c$ indicates that NE arm of the SMC is 
located at a distance $\sim~6$ kpc below the plane of the SMC main body, 
hence making it nearer to us by $\sim~6$ kpc than the main SMC plane. Several 
studies using Cepheids in the SMC found that the northeast arm of the SMC is 
nearer to us than the central bar \citep{cald86,math88,shar02}. This 
study using statistically siginificant number of RRab stars further provides 
an insight into the issue thus confirming the previous findings. The viewing 
angle parameters for the NE arm are $i= 2^{\circ}.205\pm0^{\circ}.407$ and 
$\theta_{\text{lon}} = 85^{\circ}.354\pm6^{\circ}.873$. 
Fig.~(\ref{smc_residuals}) shows the residuals after the NE arm planefit 
values were subtracted from their corresponding $z$-values. A highly symmetric 
hyperbolic paraboloid warp of very low amplitude ($\sim 0.03$ kpc) in the 
residuals is clearly discernible. Tidal effects produced by the LMC 
(Large Magellanic Cloud) and the SMC main body may be attributed to this 
warp.  Similarly, the ellipsoid fitting 
procedure is applied to the main body of the SMC which yields the values of 
$i = 0^{\circ}.178\pm0^{\circ}.091$ and $\theta_{\text{lon}} = 57^{\circ}.053\pm0^{\circ}.644$ and axes ratios as $1.000\pm0.002:1.185\pm0.004:8.298\pm0.131$. 
The presence of metallicity gradient as a function of galacto-centric distance 
of the SMC is also investigated using the empirical relation of \citet{smol05} 
with galacto-centric radius determined from inclination and position angle 
values as obtained from the $I$-band data where distance is determined from the 
\citet{cate08} empirical absolute magnitude relation. The gradient 
obtained is $-0.008\pm0.058$ dex kpc$^{-1}$. This is similar to the gradient 
of $0.00\pm0.06$ dex kpc$^{-1}$ obtained by \citet{hasc_met} 
corroborating evidence of no gradient and is contrary to the value  of 
$-0.013\pm0.007$ dex kpc$^{-1}$ obtained by \citet{kapa12} from the analysis 
of OGLE-III $V$-band light curves of $454$ RRab stars. Fewer number of data 
points and poor phase coverage in the case of $V$-band light curves of 
OGLE-III as compared to the $I$-band light curves yield unreliable estimates 
of parameters determined using them and hence should be taken with 
caution. The data in the $V$-band  also suffer from spatial resolution which 
might also result into poor estimates of viewing angles and geometric parameter 
determinations for the SMC. Fig.~\ref{hist_smol} shows cumulative distribution 
of extinction corrected RRab magnitudes $V_{0}$ for the RRab stars in the 
calibrated $V$-band as well as metallicities calculated using \citet{smol05} 
relation. The probability values obtained from the two sample 
KS tests applied on stars belonging to the NE arm (sample 1) and the SMC 
main body (sample 2) show that the CDFs of the two samples are significantly 
different at a level of $0.05$ with sample 1 nearer to us and contains more 
metal rich stars as compared to sample 2, thus confirming the fact that the 
NE arm of the SMC is nearer to us and contains more metal rich stars. 
We have also investigated the LOS depth of the sample by dividing into the 
metal poor ($[Fe/H] < \overline{[Fe/H]}-\sigma =-1.96$ dex) and metal 
rich stars ($[Fe/H] > \overline{[Fe/H]}+\sigma = -1.58$ dex). In order to find
if there is any variation in the LOS depth of the two samples belonging to 
different metallicity groups at a significance level of $0.05$, we again 
perform a two sample KS test. The two sample KS test yields $D = 0.150$ and 
$P$-value = $0.79$. The $P$-value $> 0.05$ indicates that we fail 
to reject the null hypothesis implying that there is no evidence in 
the data to suggest that the two CDFs are different. This supports the result 
by \citet{hasc_smc} that there are no variations in the 
LOS depth for different metallicity groups along the different SMC fields as 
observed by OGLE III photometric survey and is contrary to that obtained by 
\citet{kapa12}. While calculating the LOS depth, it 
has been found that the observed uncertainties in the distance measurements 
($\sigma_{\text{err}}$) of a few 
stars are greater than the observed depth ($\sigma_{\text{obs}}$). The LOS 
depth determinations for those stars have been left out. The mean LOS depth 
($\overline{\rm LOS}$) of the SMC is found to be  $4.31 \pm 0.34$ kpc. This 
value is in good agreement with the values of $4.13\pm0.27$ kpc, $4.07\pm1.68$ 
kpc and $4.2\pm0.3$ kpc found by \citet{kapa11}, \citet{smit12} and 
\citet{hasc_smc}, respectively. 

In order to see the line of sight depth variations in the SMC, we have divided 
the sample into three parts, viz., NE ($x<-0.5,~y>0.5$ kpc), central 
($-0.5<x<0.5,-0.5<y<0.5~\text{kpc})$ and SW ($x > 0.5~,y<-0.5$~\text{kpc}). 
In order to see the variation of depth among the three different 
regions of the SMC, we study their normalized LOS CDF distribution through KS 
test. A two sample KS test indicates that that LOS depths of the NE and 
SW parts of the SMC are not significantly different. On the other hand, 
KS test applied on the LOS CDF distribution of the central part and other 
parts combined (NE+SW) yields $D=0.40$ and $P$-value=$0.002$. These values 
indicate that CDF of the two distributions are different at a significance 
level of $0.05$, LOS depth of the central part being larger than rest of 
parts of the SMC. This supports the findings of \cite{subr09} and 
\citet{kapa12} that the SMC may have a central bulge.      
\section{Scale length and tidal radius of the SMC}
\label{scale}
\begin{figure}
\begin{center}
\includegraphics[width=0.46\textwidth,keepaspectratio]{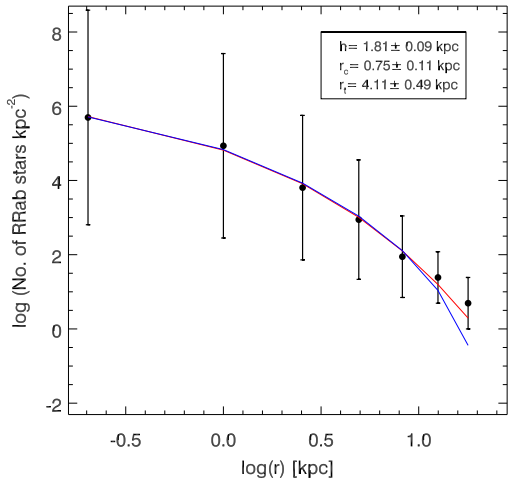}
\caption{Radial density distribution of the RRab stars in he SMC is shown as 
black circles. Error bars are due to the Poissonian distribution 
($\sigma \sim \sqrt{N}$). Red and blue solid lines show the model fits with the empirical two parameter exponential and three parameter density profiles \citep{king62}.}
\label{radial}
\end{center}
\end{figure}

Scale length is defined as the point where the number density of stars falls 
by a factor of $1/e$. Radial number density profiles of RRab stars are 
obtained by finding the radial number density of stars around the centroid 
of the galaxy in a given galacto-centric radius with increasing binsize of 
$0.5$ kpc and is shown in Fig.\ref{radial}. The number density profile is 
modeled with an exponential disk to 
find the scale length. To find the extent of the SMC in the $xy$-plane, 
radial number density distribution is fitted with King's three parameter 
profile \citep{king62}. The tidal radius gives an estimate of full extent of 
the SMC in the $xy$-plane. King's three parameter profile is described by 
\citep{king62}
\begin{equation}
n(r)=n(r_{0k}){\left(\frac{1}{[1+(r/r_{c})^2]^{\frac{1}{2}}}-\frac{1}{[1+(r_{t}/r_{c})^2]^{\frac{1}{2}}}\right)}^{2}.
\end{equation}                      
The exponential disk profile is given by
\begin{equation}
n(r)=n(r_{0e})\exp{(-\frac{r}{h})},
\end{equation}  
where $n(r_{0})$ and $h$ represent the density of RRab stars near the galactic
center and scale length, respectively. $r$ is the galacto-centric distance in 
kpc.  $r_{t}$ is called the tidal radius and represents the value of $r$ at 
which $n(r)$ falls to zero and $r_{c}$ is the core radius where the 
number density of stars falls to half its central value \citep{king62}. The 
various parameters obtained from modeling the radial density profile with 
exponential disk profile and three parameter King's profile are shown in 
Table~\ref{table_king}. The variance weighted mean values were obtained as: 
$r_{c} = 0.73\pm 0.02$ kpc, $r_{t} = 4.03 \pm 0.11$ kpc and 
$h = 1.84 \pm 0.04$ kpc. Although the values of $h$ and $r_{c}$ found here are 
comparable to the values of $1.87\pm 0.10$ kpc and $1.082\pm 0.02$ kpc, the 
value of $r_{t} = 4.03\pm0.11$ kpc is drastically different from the value of 
$7\pm 1$ kpc obtained by \citet{smit12}. Tidal radius is comparable to the  
$1\sigma$ LOS depth of $\sim 4.00$ kpc obtained in this analysis. This implies 
that extent of SMC in the $xy$-plane is comparable to the front-to-back 
distance along the line of sight and points to the possibility that RRab stars 
in the SMC are distributed more like a spheroid/slightly ellipsoid. The values 
of $r_{c}$ and $r_{t}$ imply that the concentration parameter defined as 
$c = \log{(r_{t}/r_{c})}$ \citep{king62} has a value $\sim 1.71\pm0.07$. 
\begin{table}
\begin{center}
\caption{Parameters of exponential and King's profiles for the radial number 
density distribution using the $I$-band data}
\label{table_king}
\scalebox{0.7}{
\begin{tabular}{cccccc} \\ \hline
Method&$n(r_{0e})$& $n(r_{0k})$&$h$&$r_{c}$&$r_{t}$ \\ 
& (stars kpc$^{-2}$)&(stars kpc$^{-2}$)&(kpc)&(kpc)&(kpc) \\ \hline
MVF&$844\pm73 $ & $791\pm62$ & $1.89\pm0.10$ &$0.72\pm0.10$ & $3.91\pm0.46$ \\ 
MVC&$796\pm69 $ & $746\pm60$ & $1.84\pm0.09$ &$0.72\pm0.10$ & $4.10\pm0.49$ \\
MVS&$755\pm67 $ & $711\pm58$ & $1.81\pm0.09$ &$0.75\pm0.11$ & $4.11\pm0.49$ \\
\hline
\end{tabular}
}
\end{center}
\end{table}
\section{Summary and Conclusions}
\label{summary}
In this paper, we have studied both $V$ and $I$-band light curves of RRab 
stars obtained from the OGLE-III project. Metallicities of RRab stars 
in the present sample were determined using the \citet{jurc96} relation for
both $V$ and $I$-band data. Absolute magnitudes of the RRab stars were 
determined in both ways from the Fourier parameter inter-relation of 
\citet{kova01}  and those given by \citet{cate08}. Results of the above 
analysis were further substantiated using the $I$-band data with 
\citet{smol05} metallicity relation for the $I$-band and empirical 
$M_{V}-[Fe/H]$ relation of \citet{cate08}. Also, since the $I$-band light 
curve data have very good phase coverage, we adopt the final results obtained 
from the $I$-band data only. Based on the metallicity 
of RRab stars combined with their distance distributions in the SMC, the 
following conclusions are drawn:
\begin{enumerate}
\item RR Lyrae stars in NE arm are nearer to us than the $xy$-plane of the 
SMC main body by a distance of $\sim6$ kpc. Apart from that, the NE arm 
contains more metal rich stars as seen from metallicity distributions of 
these two groups using the two sample KS-statistics. Using the age, metal 
abundance and positional data of populous cluster using Cepheids in the SMC, 
\citet{crow01} found that the eastern side of the SMC containing Cepheids is 
nearer to us and contains more metal rich stars but warrants further study. 
The investigation done in this study bears out this fact. The existence of an 
isolated NE arm of the SMC was further corroborated using the \citet{schl98} 
reddening map. However, it indicates that the NE arm is nearer to us by $\sim 11$ 
kpc than the galactic plane of the SMC main body. The \citet{schl98} maps use 
$IRAS$ 
DIRBE data on the far-IR sky emission and estimate the extinction from the dust 
properties directly. Nonethless, the Schlegel maps are highly uncertain in the 
inner region of the SMC because their temperature structures were not 
sufficiently resolved by DIRBE \citep{schl98,pess06}. The choice of different
search radius as well as different reddening maps may have some effect on the 
final result of this paper. Availability  of accurate and highly precise 
extinction measurements covering the entire region of the SMC will help 
enhance our understanding of its structure in great details.              
\item Modeling the observed population of RRab stars in the SMC by a 
triaxial ellipsoid, we estimated the axes ratio, inclination of longest axis 
with the line of sight ($i$) and the position angle of line of nodes 
$(\theta_{\text{lon}})$ on the sky from the analysis of the entire sample of 
RRab stars. Axes ratios of the galaxy were obtained as 
$1.00\pm 0.000:1.310\pm 0.029:8.269\pm0.934$ with $i = 2^{\circ}.265\pm 0^{\circ}.784, ~\theta_{\text{lon}}=74^{\circ}.307\pm 0^{\circ}.509$ from the variance
weighted $I$-band determinations. Using the $12$ populous clusters in the SMC, 
\citet{crow01} obtained the axes ratios approximately as 
$1:2:4$ modeling the SMC as a triaxial galaxy with declination, right 
ascension and line of sight depth as the three axes. On the other hand, using 
the same OGLE-III SMC RRab dataset, \citet{smit12} obtained the axes ratios, 
inclination of the longest axis with the line of sight ($i$) and the position
angle of the projection of the ellipsoid on the plane of the sky 
($\theta_{\text{lon}}$) as $1.00:1.30:6.47,~0^{\circ}.4,~74^{\circ}.4$, 
respectively, with no errors in the parameters quoted. Whereas, 
\citet{hasc_smc} found an inclination angle of $7^{\circ}\pm 15^{\circ}$  and 
a position angle of $83^{\circ}\pm 21^{\circ}$ from $1494$ OGLE-III RR Lyrae 
stars using only the $I$-band data. These values are quite consistent with 
those obtained in the present study using distance distributions determined 
from the $M_{V}-[Fe/H]$ and $M_{V}$-Fourier parameters relations.
\item NE arm of the SMC resembles a three-dimensional plane with the plane
slightly warped in a hyperbolic paraboloid, while the the SMC main body is 
distributed like an ellipsoid. The NE arm has been fitted by a plane equation 
to determine its viewing angles. The SMC main body has been fitted with an 
ellipsoid obtained from the principal axis transformation of the moment of 
inertia tensor constructed from the $(x,y,z)$ coordinates. Analyses indicate
that the NE arm is slightly misaligned with the SMC main body, with a highly 
symmetric warp of low amplitude ($\sim 0.03$ kpc). The warp may have been 
caused by the tidal forces exerted on it by the main SMC body and the LMC. 
Axes ratios of the SMC main body were obtained as $1.00\pm 0.000:1.185\pm 0.001:9.411\pm0.860$ with $i = 0^{\circ}.507\pm 0^{\circ}.287, ~\theta_{\text{lon}}=55^{\circ}.966\pm 0^{\circ}.814$ from the variance weighted $I$-band 
determinations. On the other hand, viewing angle parameters of the NE arm are 
found to be $i = 2^{\circ}.244\pm 0^{\circ}.024$ and $
\theta_{\text{lon}} =85^{\circ}.541 \pm 0^{\circ}.332$ from the variance 
weighted $I$-band determinations.    
\item Combining metallicities with spatial distribution of these tracers, no 
radial metallicity gradient in the SMC has been detected. Dividing the entire 
sample into three parts: northeastern (NE), central and southwestern (SW), we 
found that the central part has a significantly larger line of sight depth as 
compared to rest of the parts, indicating that the SMC may have a bulge. 
Independent analysis done in this work exploring different empirical relations
and photometric bands provides substantial evidence to the findings of 
\citet{subr09} and \citet{kapa12} of SMC having a central bulge.        
\item Radial number density of RRab stars in the $I$-band were modeled by
an exponential disk and three parameter \citet{king62} profiles. 
Following values were obtained from the modeling: $h = 1.84\pm0.04$ kpc, 
$r_{c} = 0.73\pm 0.02$ kpc, $r_{t} = 4.03\pm 0.11$ kpc and $c = 1.71\pm 0.07$.  \end{enumerate}          
\section*{Acknowledgments}  
The authors thank the OGLE team for making their invaluable variable star 
data publicly available. SD thanks Department of Science \& Technology (DST), 
Govt. of India for support through a research grant D.O No. 
$\text{SB/FTP/PS-029/2013}$ under the  Fast Track Scheme for Young Scientists. 
HPS and SMK acknowledge Indo-US Science \& Technology Forum (IUSSTF) for 
support. SD and SK also acknowledge support through funding in the DU 
innovation scheme by University of Delhi. The authors acknowledge helpful 
discussions with Chow Choong Ngeow on extinction determinations. The use of 
arxiv.org/archive/astro-ph and NASA ADS databases is highly acknowledged. 
The authors thank the anonymous referee for all the insightful comments 
and suggestions that significantly improved the paper.
\bibliographystyle{mn2e}
\bibliography{deb}
\end{document}